\documentclass[twocolumn, aps, longbibliography]{revtex4-1}

 \pdfoutput=1
\usepackage[utf8]{inputenc}
\usepackage{graphicx}
\usepackage{color}
\usepackage{epsfig}
\usepackage{subfigure}
\usepackage{natbib}
\usepackage{array}
\usepackage{amsmath}
\usepackage{ulem}

\usepackage{tikz}

\makeatletter

\begin{document}
\newcolumntype{M}{>{$}c<{$}}


\title{Optimum surface-passivation schemes for near-surface spin defects in silicon carbide}

\author{Cyrille Armel Sayou Ngomsi}
\author{Tamanna Joshi}
\author{Pratibha Dev}
\affiliation{Department of Physics and Astronomy, Howard University, Washington, D.C. 20059, USA}
\date{\today}

\begin{abstract}

 Spin-active defects in silicon carbide (SiC) are promising quantum light sources for realizing scalable quantum technologies. In different applications, these photoluminescent defects are often placed in a nanostructured host or close to surfaces in order to enhance the signal from the defects.  However, proximity to the surface not only modifies frequencies of the quantum emission from the defect, but also adversely affects their photo-stability, resulting in blinking and/or photobleaching of the defect.  These effects can be ameliorated by passivating surfaces with optimal adsorbates.  
In this work, we explore different passivation schemes using density functional theory-based calculations.  We show that a uniform surface passivation with either hydrogen or with mixed hydrogen/hydroxyl groups completely removes surface states from the SiC band gap, restoring the optical properties of the defects.  

\end{abstract}

\maketitle

\section{Introduction}



Luminescent point defects in different polytypes of silicon carbide (SiC) are being explored as quantum bits (qubits) due to their addressable spins~\cite{falk2013polytype,widmann2015coherentRT,Seo2016,nagy2019highfidelity}, long spin coherence times~\cite{Seo2016,Christle2015,Carter2015,nagy2019highfidelity,widmann2015coherentRT}, and spectral stability \cite{Udvarhelyi2019,babin2021nanofabricated}. 
In addition, different defect-based quantum emitters in SiC display near-telecom range emissions~\cite{Christle2015,Fuchs2015}, resulting in relatively low propagation losses. SiC itself is cheap and commercially available as a semiconductor-grade material with well-developed fabrication protocols, giving an edge to the quantum emitters in SiC over those in diamond~\cite{balasubramanian2008,KaxirasNV,stanwix2010,tesfaye2011,doherty2013nitrogen,bhandari2021} for creating scalable quantum technologies.

In order to employ point defects in quantum computing and sensing applications, one often places them in close proximity to surfaces or within nanostructures, such as photonic crystal cavities fabricated in SiC hosts~\cite{gadalla2021enhanced,majety2021quantum,crookEHu2020purcell, bracherEHu2017selective} or in arrays of nanopillars~\cite{nagy2018quantumdichroic,lukinRadulaski20204h,radulaski2017scalable}. Such arrangements, while advantageous, can also introduce complications due to surface and quantum confinement effects that can change quantum emitter properties in unpredictable ways.  Over the last few years, this realization has resulted in several experimental works on near-surface defects in different semiconductors hosts of qubits~\cite{lukinRadulaski20204h,YuanDeLeon2020Charge,sangtawesinDeLeon2019origins,nanostr_VSi_expt_2020,neethirajan2023}. However, on the theoretical side, most \textit{ab-initio} works concentrate on defect properties in bulk crystals, with a very few exceptions that have considered deep-level defects in nanostructured hosts~\cite{Kaviani2014,LiGali2019,Lofgren_2019,Joshi2022}. 
In particular, a recent theoretical work by Joshi \textit{et al.}~\cite{Joshi2022} started to address the aforementioned knowledge gap 
by considering a negatively charged silicon monovacancy ($\mathrm{V_{Si}^{-1}}$) in a 2H-SiC nanowire (NW).  
Joshi \textit{et al.} showed that different finite size effects, such as the naturally-present strain in the unpassivated NW and the surface states at the valence and conduction band edges, profoundly affect properties of the near-surface silicon monovacancies.  As a result, the frequency of quantum emission from $\mathrm{V_{Si}^{-1}}$ in a NW changes dramatically relative to that in bulk. 
They further showed that these finite-size effects can lead to deterioration of the device performance due to the defect's charge-state conversion from the bright negatively-charged state to the dark neutral state ($\mathrm{V_{Si}^{0}}$) upon photoexcitation. 
The charge-state conversion itself is facilitated by spatial and energetic overlap between the defect and surface states, resulting in blinking or photobleaching of the defect.
\begin{figure*}[htb]
    \centering
    \includegraphics[width = 0.7\textwidth]{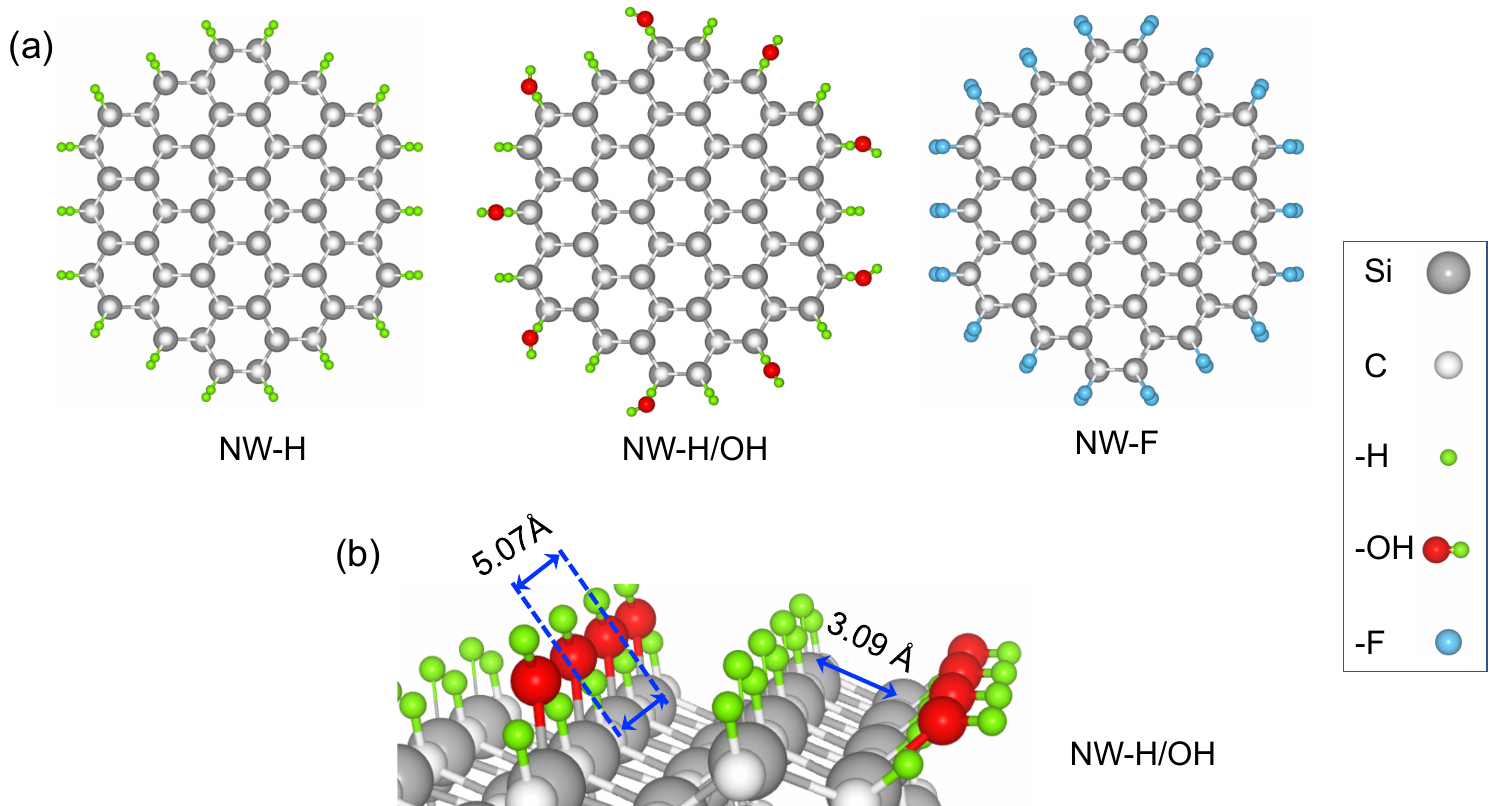}
      \vspace{-6pt}
    \caption{Passivated silicon carbide nanowire (NW). (a) From left to right: top view of a NW with hydrogen (-H), mixed hydrogen/hydroxyl group (-H/OH) and fluorine (-F) passivation of three-fold coordinated silicon (Si) and carbon (C) atoms on the surface. (b) A close-up of the NW surface with the -H/OH passivation. The particular scheme chosen here reduces steric repulsion between the hydroxyl groups. This is ensured by alternating rows of -H and -OH groups on nearest-neighbor silicon atoms at the edges (separated by about 3.09\,\AA). All carbons are passivated with hydrogens. This is just one such structure that can be created and is used as a ``model" for mixed -H/OH passivation.}
    \label{fig:fig1}
\end{figure*}


While the aforementioned work showed how the finite size effects are detrimental to the properties of near-surface defects in nanostructured hosts, Joshi \textit{et al.} did not consider potential passivation schemes that can ameliorate these effects.  In this density functional theory (DFT)-based work, we have addressed the issue of charge-state instability of the near-surface $\mathrm{V_{Si}^{-1}}$ by identifying different chemical terminations for the surfaces of the nanostructured SiC that effectively remove the surface states from the band gap of the nanostructure.  In particular, we show that surface passivation with either hydrogen (-H) or mixed hydrogen/hydroxyl groups (-H/OH) produce the desired effects, while fluorination (-F) should be avoided.  In both of the preferable chemical passivation schemes, we find near-perfect restoration of the frequency of quantum emission and the removal of surface states from the valence and conduction band edges, signifying  increased photostability of the defect-based quantum emitter.  



\section{Computational details}

We employed the QUANTUM ESPRESSO package~\cite{QE-2009,QE-2017} to perform spin-polarized, DFT-based calculations. 
Ultrasoft pseudopotentials~\cite{USPP_vanderbilt1990} were used to describe electron-ion interactions, allowing for a low energy cut-off and hence, a relatively small plane-waves basis set for representing the wave functions. We found that the energy cutoffs of 50\,Ry for expanding wave functions and 350\,Ry for charge densities were sufficient to ensure convergence.  Ionic relaxation was performed for all structures until the forces on the atoms were smaller than $10^{-3}$\,Ry/a.u.
In our calculations, exchange-correlation effects were included within the generalized gradient potential approximation (GGA)~\cite{GGA} of Perdew, Burke, and Ernzerhof (PBE)~\cite{PBE}.  Although the semilocal PBE functional is known to underestimate band gaps~\cite{PBE_bandgap_issue}, we chose to use it instead of computationally expensive hybrid functionals, such as HSE06~\cite{HSE03, HSE06}.  We made this choice because: (i) it made calculations feasible for the large structures used in the present study and (ii) in our work, we are interested in determining the quantum emission energies of the near-surface defects within nanostructures (without and with surface passivations) relative to the quantum emission energy of the defect in bulk 2H-SiC. Our emphasis on \textit{energy-differences} rather than their absolute values minimizes the error in our predictions and allows us to use the PBE functional without reducing the accuracy of our analysis. That PBE may become a more judicious choice over the prohibitively-expensive HSE06 functional can be seen in DFT calculations within the two functionals that yielded very similar adiabatic potential energy surfaces for a quantum emitter in hexagonal boron nitride~\cite{noh2018_PBE,li2020_hse}. 


In order to facilitate a direct comparison with the results obtained by Joshi \textit{et al.}~\cite{Joshi2022}, we have adopted the bulk and NW structures as described in their work.  Hence, we chose to create our nanostructure using the 2H-SiC polytype. 
The selection of the 2H-polytype allowed us to create a stoichiometric nanostructure with no magnetic moment on any atom, including those on the NW surface. This, in turn, ensured that the spin of $\mathrm{V_{Si}^{-1}}$, once it was introduced, did not interact with other surface-related spins. 
Hence, the choice of the 2H-SiC polytype helped to isolate and emphasize different surface and quantum confinement effects that modify properties of the quantum emitters over any other spurious effects. The unpassivated (``bare"), as-created nanowire was periodic in the $[0001]$-direction (c-axis) and consisted of $216$ atoms. The equilibrium diameter of the bare NW was  $15.43$\,\AA. The interactions between the periodic images of the nanowire in the lateral direction were minimized by adding a vacuum of about 12\,\AA{} around the NW in the plane perpendicular to the c-axis.  All calculations for NWs were performed using a gamma-centered k-point grid of $1\times1\times6$, created according to the Monkhorst-Pack scheme~\cite{Monkhorst}.  For bulk 2H-SiC, we used an  $8\times8\times2$ supercell, consisting of 512-atoms.  Although this was a large supercell, a $\Gamma$-centered $2\times 2 \times 6$ k-grid was used, which has been found to be necessary to obtain converged results for this defect~\cite{Carter2015,soykal2016silicon,economou2016spin,Joshi2022}. 

Starting from the pristine, as-created NW, we passivated the dangling bonds on the NW surface with simple adsorbates (-H, mixed -H/OH or -F). Figure~\ref{fig:fig1}(a) shows the equilibrium geometries for the defect-free, passivated NWs. Single covalent bonds are formed between the three-fold coordinated Si and C atoms on the surface and the adsorbates. It should be pointed out that there are a number of possible structural models for the mixed -H/OH passivation, such as: (i) the ``Si-OH\,+\,C-H" structure, where the surface silicon-atoms (carbon-atoms) are passivated with hydroxyl groups (hydrogens), and (ii) the ``Si-H\,+\,C-OH" structure, where the surface silicon atoms (carbon atoms) are passivated with hydrogen atoms (hydroxyl group).  Since the former surface passivation scheme (i.e. Si-OH\,+\,C-H) was recently shown to be energetically more favorable~\cite{passivated_SiC_NW_YaHui2020}, we chose to adopt it, but with a modification to further reduce steric repulsion.  Since the silicon atoms on the surface can be as close as 3.09\,\AA, a uniform passivation of all under-coordinated Si atoms on the surface with the -OH group can result in the crowding of hydroxyl groups. In our modified scheme for -H/OH passivation, all under-coordinated C atoms were passivated with hydrogens, while it was ensured that the Si atoms on the surface that were closer to each other have alternating -H and -OH groups, as shown in Fig.~\ref{fig:fig1}(b).

Effects of surface passivation on the structural, electronic, and optical properties of the near-surface defects were studied by creating a $\mathrm{V_{Si}^{-1}}$ defect in the passivated 2H-SiC NWs. In order to study if the adsorbates affect the doping conditions under which the negative charge state of the defect is stable, we calculated the defect formation energies as a function of the electronic chemical potential (i.e. the Fermi energy, $E_F$). The formation energy, $\Delta E_{form}$, of a defect $X$ (here, $\mathrm{V_{Si}}$) in charge state $q$ is defined as~\cite{tanaka2003}:
\begin{equation}\label{formation_energy}
\begin{aligned}
   \Delta E_{form}(X; q) =  & E_{Total}(X; q) - E_{Total}(ideal,0) \\
  & + \mu_{\textrm{Si}}  + q \left[E_{VBM}(X) + E_F\right] \\ \\
  \end{aligned}
\end{equation}
\noindent  where $\mu_{\textrm{Si}}$ is the chemical potential of Si, accounting for the removal of a Si-atom from the supercell to create $\mathrm{V_{Si}}$, and $E_{Total}(X; q)$ [$E_{Total}(ideal,0)$] is the total energy of the defective [pristine/ideal] supercell.  The valence-band maximum, $E_{VBM} (X)$, of the defective supercell is calculated using: 
\begin{equation}\label{vbm_align}
\begin{aligned}
E_{VBM} (X) = E_{VBM} (ideal) + V_{av} (X)- V_{av} (ideal) 
  \end{aligned}
\end{equation}
\noindent Here, $E_{VBM}(ideal)$ is obtained from the difference in the total energies of the neutral and positively-charged ideal/perfect supercells [i.e. $E_{VBM}(ideal)=E_{Total} (ideal; 0) - E_{Total} (ideal; +1)$].
\noindent In Eq.~\ref{vbm_align}, since we are using $E_{VBM} (ideal)$ to determine $E_{VBM} (X)$, we need to ensure that  the band structures of the defective and ideal structures line up. This is done by assuming that far away from the defect, the potential in the defective structure should be similar to that in the ideal. Hence, the difference between the averaged potentials ($V_{av}$) between the perfect and defective supercells provides the offset between $E_{VBM} (ideal)$ and $E_{VBM} (X)$.

In order to determine how passivation affects the optical properties of the near-surface $\mathrm{V_{Si}^{-1}}$, we calculated the zero-phonon line (ZPL) using the $\Delta$SCF method~\cite{KaxirasNV}, which constrains the occupation of the defect states to emulate the photo-excitation process. The $\Delta$SCF method provides a computationally inexpensive means of determining ZPL values and 
has been successfully employed to study excitations between spatially-localized defect states of deep-level defects in different wide band gap semiconductors~\cite{PDEV_hBN_2020,Joshi2022,narayanan2023}. In order to obtain the ZPL of the defect using the $\Delta$SCF method, one starts from the ground state equilibrium structure given by generalized coordinate, $Q_{ground}$, and the ground state electronic configuration of the system, corresponding to point A of the Franck-Condon picture shown in Fig.~\ref{fig:fig2}(a). The vertical excitation process is then emulated by constraining electronic occupation to that of the excited state. This is achieved by emptying a filled defect state and promoting it to an empty defect state of the same spin, while leaving the ionic position unchanged [point B in Fig.~\ref{fig:fig2}(a)]. The point C in Fig.~\ref{fig:fig2}(a) is obtained by relaxing the structure, while constraining the electronic occupation to that for the excited state, emulating ionic relaxation in response to the changed charge density distribution. Lastly, the point D is obtained by constraining the ionic positions to those obtained in previous step ($Q_{excited}$), while requiring a ground state electronic configuration. A difference in total energies of the system at points A and C gives the ZPL.

\section{Results}

\subsection{Silicon vacancy in bulk SiC and bare SiC NW}
\begin{figure*}[htb]
    \centering
    \includegraphics[width = 0.8\textwidth]{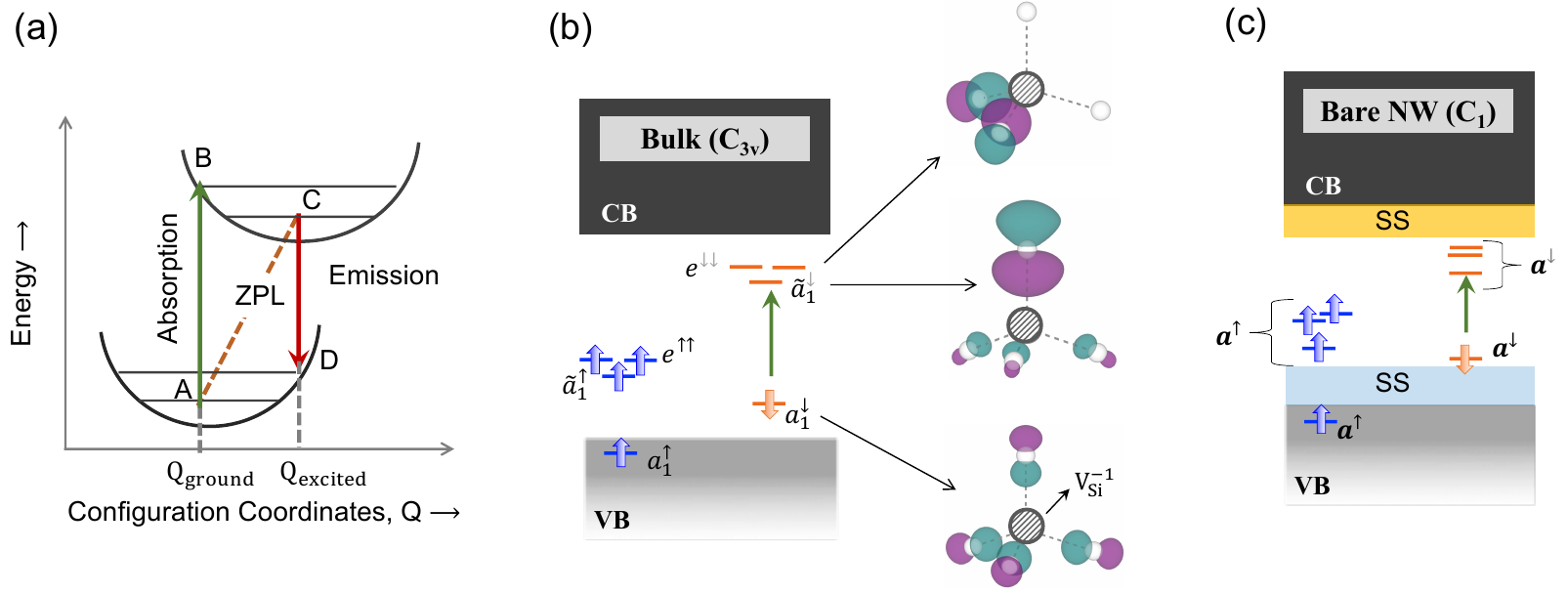}
      \vspace{-6pt}
    \caption{ (a) Franck-Condon picture onto which the $\Delta$SCF results are mapped to emulate the photoexcitation process. (b)  Schematic single particle level diagram (not to scale), showing the defect states introduced by $\mathrm{V}_{\mathrm{Si}}^{-1}$ in 2H-SiC bulk, where the structure has a $C_{3v}$-point group symmetry. The spin-split defect states are labelled by their group theoretical representations, with black (grey) arrows in the superscripts used to emphasize that the states are filled (empty).  The green vertical arrow in the energy level diagram indicates the lowest energy photo-excitation in bulk. In the charge density plots on the right for the optically-active, minority-spin (spin-down) defect states, only the carbon atoms surrounding $\mathrm{V}_{\mathrm{Si}}^{-1}$ are shown for clarity.  Teal (mauve) color corresponds to positive (negative) isovalues, highlighting the bonding and/or antibonding character of the defect states. (c) Schematic energy-level diagram (not to scale), showing the changes in the electronic structure of $\mathrm{V}_{\mathrm{Si}}^{-1}$ when created in an unpassivated/bare NW. 
 Here, VB, CB, and SS refer to the valence band, conduction band and surface states, respectively.}
    \label{fig:fig2}
\end{figure*}
Before discussing our results for the passivated NW, we briefly describe those for bulk and the bare NW to highlight the importance of finding appropriate surface coverages. A greater in-depth exposition can be found in Ref.~\cite{Joshi2022}.  In different hexagonal polytypes of bulk SiC, the negatively charged monovacancy, $\text{V}_{\text{Si}}^{-1}$, is a spin-3/2 defect that has a $C_{3v}$ point group symmetry~\cite{VSiq0_Spin1_Torpo1999,Mizuochi2002,janzen2009silicon,Carter2015,soykal2016silicon,economou2016spin,Joshi2022}.  The dangling sigma bonds of the surrounding carbon atoms are derived from their $2s$- and $2p$-orbitals, which are known to result in spatially localized defect states~\cite{Dev_PRL_DeepDefects_2008, Dev_PRB_DeepDefects_2010, Dev_PRB_NW_2010, PDEV_hBN_2020,Joshi2022,narayanan2023}.  The single-electron molecular orbitals for the defect states can be obtained from the dangling sigma bonds, resulting in two singlet states [herein denoted as $a_{1}$ and $\tilde{a}_{1}$]  and a doublet [$e=\{e_X, e_Y\}$]. Figure~\ref{fig:fig2}(b) is a schematic diagram showing the spin-split single particle energy levels (not to scale). 
The distribution of five electrons left behind by $\text{V}_{\text{Si}}^{-1}$ in these states results in an excess of three electrons in the majority spin channel as compared to the minority spin channel [Fig.~\ref{fig:fig2}(b)].  In the same figure, we have also shown the charge density plots for the optically-active defect states (minority-spin) that are involved in the photoexcitation process.  
For $\text{V}_{\text{Si}}^{-1}$ in bulk 2H-SiC, the lowest energy optical excitation corresponds to $a_{1}^{\uparrow\downarrow}\tilde{a}_{1}^{\uparrow}e^{\uparrow\uparrow} \rightarrow a_{1}^{\uparrow}\tilde{a}_{1}^{\uparrow\downarrow}e^{\uparrow\uparrow}$, with a reported ZPL value of 1.19\,eV~\cite{Joshi2022}.  

In contrast to the ZPL of $\text{V}_{\text{Si}}^{-1}$ in 2H-SiC bulk, a much lower ZPL of 0.87\,eV was reported~\cite{Joshi2022} for a $\text{V}_{\text{Si}}^{-1}$ defect at a near-surface site in the interior of an unpassivated/bare NW. Figure~\ref{fig:fig2}(c) is a schematic energy-level diagram (not to scale), showing the changes in the electronic structure of $\mathrm{V}_{\mathrm{Si}}^{-1}$ in the bare NW.  In the figure, VB, CB and SS refer to the valence band, conduction band and, the surface states that are introduced at the band edges, respectively. 
Out of different finite size effects, Joshi \textit{et al.} showed that the most significant factor contributing to the 325.6 meV reduction in the ZPL energy of the defect in the bare NW as compared to the bulk is the hybridization between the surface states and the defect states. The hybridization is made possible by the energetic and spatial overlap between the defect and the surface states. One consequence of this is the narrowing of the gap between the frontier orbitals (i.e. the highest-occupied and lowest-unoccupied defect states), from 1287.7 meV in bulk to 938.3 meV for a near-surface defect in the bare NW.
The energetic and spatial proximity of the surface states to the defect states not only tunes the emission frequencies, but also has detrimental effects on the charge-state stability of a monovacancy, converting it from the bright $\mathrm{V}_{\mathrm{Si}}^{-1}$ to the dark neutral defect ($\mathrm{V}_{\mathrm{Si}}^{0}$). This occurs when the surface states either temporarily trap the photoexcited electron, resulting in the blinking of the defects or permanently remove the photoexcited electron resulting in photobleaching.  In what follows, we show that appropriate passivation of the surfaces of the nanostructured hosts can counter the unpredictable ZPL changes and improve photostability of the near surface defects.   


 \subsection{Defect-free passivated NW}

\begin{figure*}[htb]
    \centering
    \includegraphics[width = 0.72\textwidth]{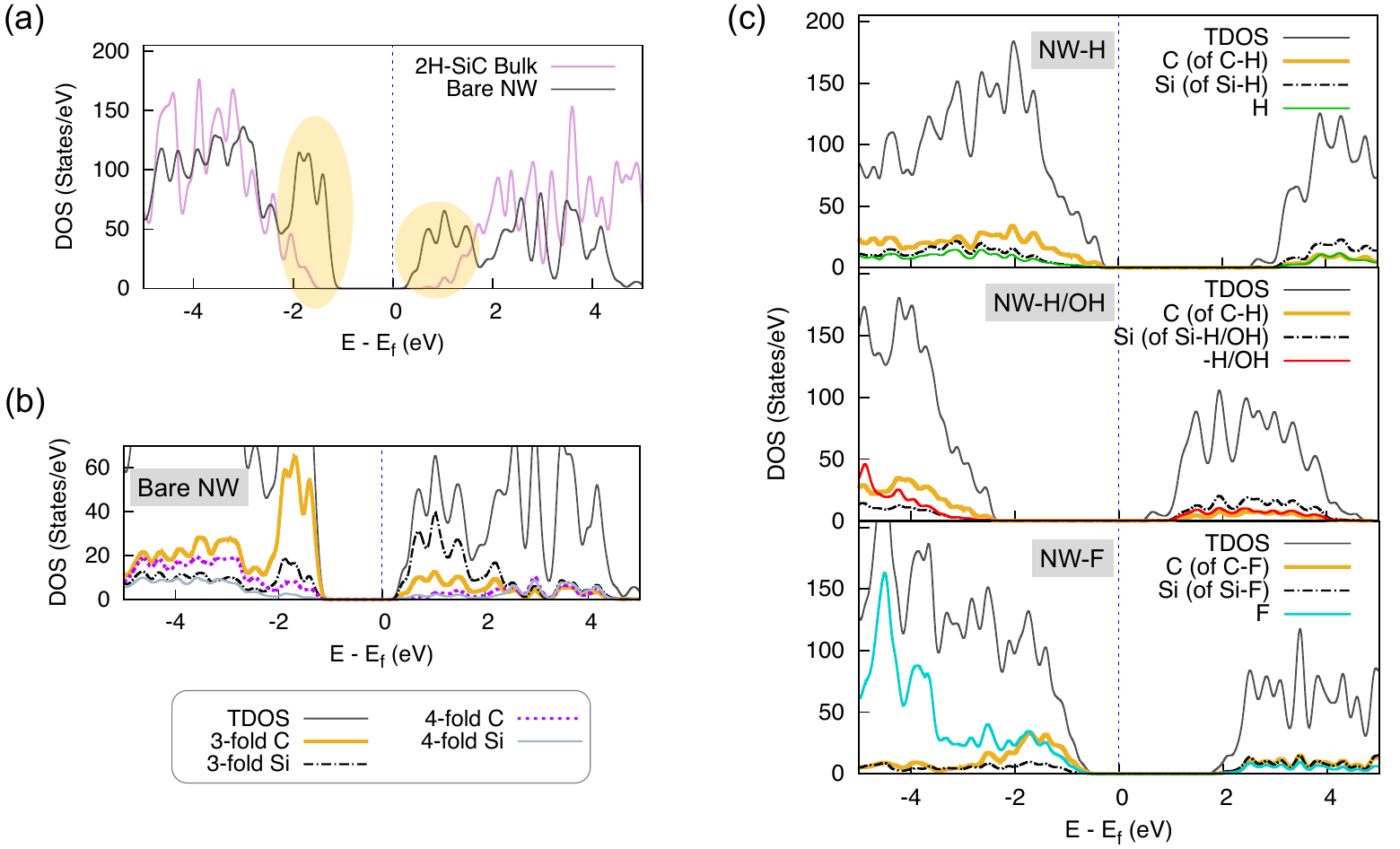}
      \vspace{-6pt}
    \caption{Electronic structure properties of defect-free 2H-SiC NWs. (a) Density of states (DOS) for unpassivated/bare NW, along with the  DOS of perfect 2H-SiC bulk, showing surface states (highlighted) at the band edges of the NW. (b) Total DOS (TDOS) and DOS projected onto the three-fold and four-fold coordinated carbons and silicons on the surface of the bare nanowire. 
 (c) From top to bottom: total and projected DOS plots for NW-H, NW-H/OH and NW-F.  The DOS projected onto the carbon and silicon atoms, which have now been passivated with the adsorbates, are shown with solid orange and dash-dotted black lines in all three plots.  Also shown is the DOS contributed by the adsorbate. The surface states are completely removed by hydrogenation and mixed -H/OH adsorption.
    The bottom plot for NW-F shows carbon-derived states at the valence band edge that are reminiscent of the surface states in the bare NW.  
    }
    \label{fig:fig3}
\end{figure*}

In their work on a pristine 2H-SiC NW, Joshi \textit{et al.}~\cite{Joshi2022} reported a considerable contraction of bonds between atoms at the under-coordinated sites on the surface of the bare NW.  This skin-bond contraction is a characteristic effect found in very small nanostructures~\cite{Huang2008SkinEffect} and aids in a reduction of the surface energy via strengthening of the remaining bonds at the undercoordinated surface sites. 
Once the under-coordinated sites on the NW surface are passivated, the structural reconstruction of bonds at these surface sites, which still have a different chemical environment than the underlying bulk, is much smaller as compared to that in the bare NW.  As a consequence, the structural changes that propagate to the interior of the passivated NW are even smaller. For example, upon hydrogenation, the Si-C bond lengths at a hydrogenated surface site contract by 1.21\% and 0.34\% in the axial- and basal-directions, respectively as compared to the bulk. These are much smaller bond contractions when compared to those reported for the surface sites of a bare NW, which showed 8.5\% and 3.20\% contraction along the axial direction and basal plane, respectively~\cite{Joshi2022}.  
For H-passivation, the interior sites now show a bond-length elongation of 0.06\% and 0.18\% in the axial- and basal-directions, respectively, relative to the bulk values. Similarly, for the mixed -H/OH passivation, the interior sites now show, on average, a bond-length elongation of about 0.05\% and 0.21\% in the axial- and basal-directions, respectively. In the case of fluorination, the overall strain in the system is larger, with the Si-C bonds showing an elongation of 0.11\% along the axial-direction and an average elongation of 0.38\% for Si-C bonds with carbons in the basal-plane.  Hence, our results show that the local strains close to the surfaces depend on the surface chemistries, and we expect them to play a role in modifying the electronic and optical properties of the near-surface defects.

 \begin{figure*}[htb]
    \centering
    \includegraphics[width = 0.75\textwidth]{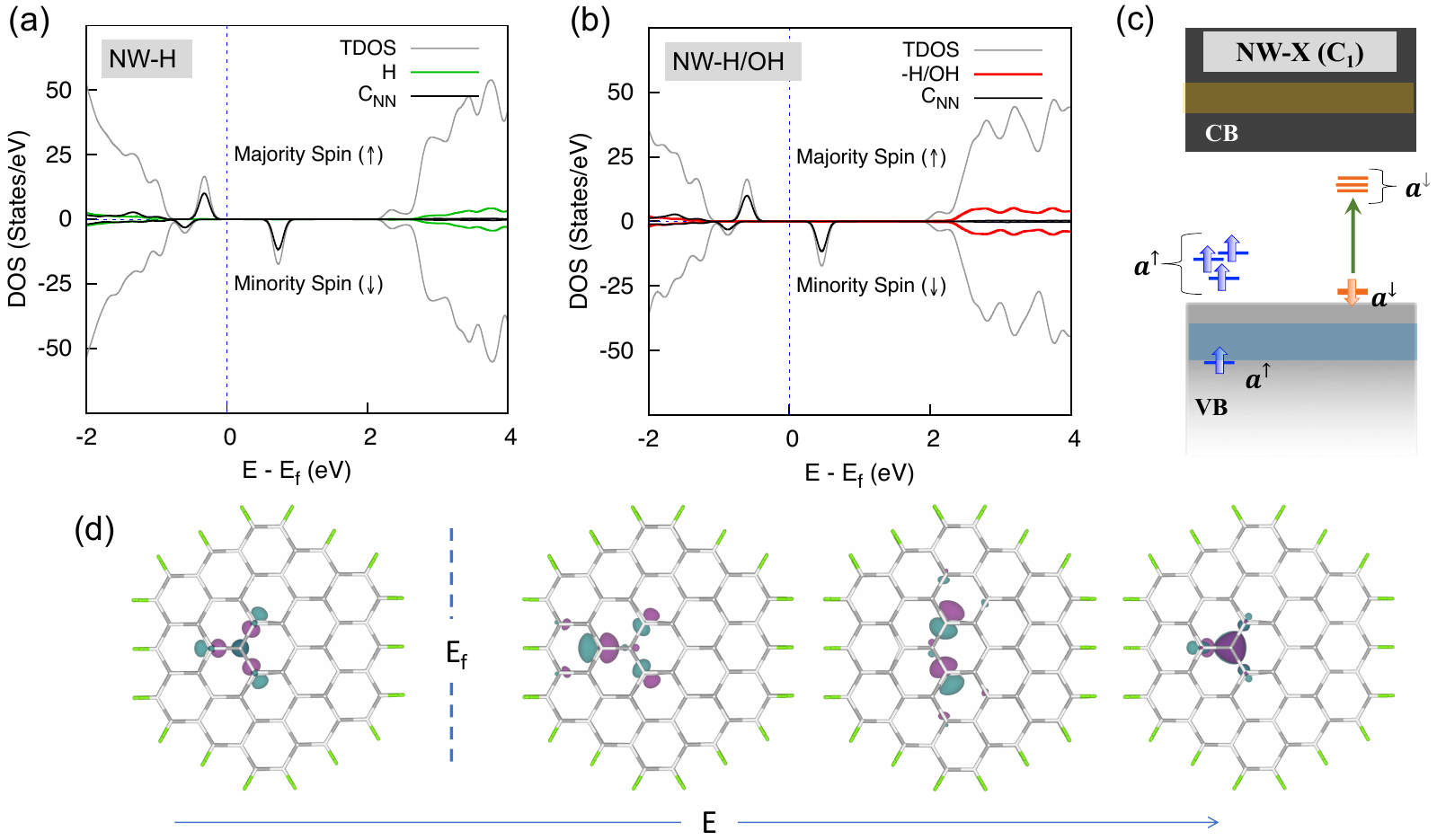}
      \vspace{-6pt}
    \caption{Hydrogen and mixed hydrogen/hydroxyl group passivated 2H-SiC NW with a $\mathrm{V_{Si}^{-1}}$ defect. (a) Total DOS (TDOS) in gray and the DOS contributed by the four nearest neighboring carbon atoms ($\mathrm{C_{NN}}$) surrounding the defect (in black). Also shown in green is the DOS projected onto the hydrogen-atoms. 
 (b) TDOS and projected DOS for NW-H/OH, showing similar behavior to NW-H.  (c) Schematic energy-level diagram (not to scale) summarizing the electronic structure for NW-X (X= -H or -H/OH), highlighting how the electronic structure properties of a defect in the passivated NW differ from those for the bare NW.  
 The dangling bond/surface states that previously occupied the band gap in the bare NW are now passivated and pushed deeper into VB and CB (presented as colored bands in VB and CB). (d) The charge density plots of the optically-active empty and filled state for NW-H. 
 Teal (mauve) color corresponds to positive (negative) isovalues, highlighting the bonding and/or antibonding character of the defect states.}
    \label{fig:fig4}
\end{figure*}

In addition to influencing the structural properties at the near surface sites, different surface chemistries also modify the electronic structure properties. Figure~\ref{fig:fig3}(a) shows the density of states (DOS) of the bare NW, along with the DOS of bulk 2H-SiC for comparison.  Although, one expects a larger band gap in the nanostructure due to quantum confinement effects as compared to the band gap (2.31\,eV) of bulk 2H-SiC, the bare NW actually has a smaller band gap of 1.73\,eV, as seen in Fig.~\ref{fig:fig3}(a). The smaller band gap of the NW is a consequence of states at the valence and conduction band edges of the NW [see the highlighted states in Fig.~\ref{fig:fig3}(a)], which are absent in the DOS of the bulk.  To understand the origin of the states at the band edges, we have plotted the DOS projected onto the three-fold coordinated C (solid orange line) and Si atoms (dash-dotted black line) in Fig.~\ref{fig:fig3}(b), showing that the main contributions to these states comes from the undercoordinated atoms at the surface. On the other hand, DOS contributions from the four-fold coordinated carbon (dotted pink line) and silicon atoms (solid blue-grey line) on the surfaces are much smaller, as seen in Fig.~\ref{fig:fig3}(b).
Figure~\ref{fig:fig3}(c) is a plot of the total and projected DOS of the passivated NWs in which different adsorbates are used to passivate the dangling bonds of the under-coordinated Si and C atoms.  The DOS projected onto the carbon and silicon atoms, which were previously under-coordinated and have now been passivated with the adsorbates, are shown with solid orange and dash-dotted black lines in all three sub-plots.  
The topmost and the middle panel in Fig.~\ref{fig:fig3}(c) shows that upon passivation with -H and mixed -H/OH, the surface states are completely removed from the band gap, resulting in the observed larger gaps of 3.02\,eV and 3.05\,eV, respectively.  On the other hand, the bottom plot for NW-F shows C-derived states at the valence band edge that are reminiscent of the surface states in the bare NW.  As a result, NW-F has a band gap of 2.54\,eV, which is smaller than those for NW-H and NW-H/OH. 

One may suspect that the smaller band gap of NW-F [see Fig.~\ref{fig:fig3}(c)] might itself be mostly due to the well-known band gap error of DFT within the PBE approximation. In turn, such an underestimation of band gap can then result in the incorrect placement of the fluorine and carbon-derived states of NW-F at the valence band edge.  
However, we note that the band gap error of DFT within the PBE approximation should similarly affect the electronic structure properties of NW-H and/or NW-H/OH. But neither of these two structures display the features present in the DOS of NW-F, indicating that our results for NW-F are not an outcome of the band gap error. In fact, we attribute our NW-F results to the  chemistry at its surface. Fluorine is the most electronegative element. As a consequence, $\mathrm{C-F}$ and $\mathrm{Si-F}$ bonds are highly ionic in nature as compared to, for example, the $\mathrm{C-H}$ and  $\mathrm{Si-H}$ bonds at the NW-H surface.  This results in very different bonding and charge distribution at the NW-F surface. 
 The differences in the nature of bonds can be quantified by  determining their fractional ionic character.  As defined by Pauling, the fractional ionic character, is given by: $f_{i}(A-B)=1-exp[-(\mathrm{EN}_{A} -\mathrm{EN}_{B})^{2}/4]$, where $EN_{A}$ and $EN_{B}$ are the electronegativities of elements A and B, respectively~\cite{Gamble1974,ML_Pankaj2022}.  Pauling's formula yields large fractional iconicity values of $f_{i}(\mathrm{Si-F})=66.1\%$  and  $f_{i}(\mathrm{C-F})=40.0\%$ for bonds at the NW-F surface. For comparison, we get $f_{i}(\mathrm{Si-H})=2.2\%$ and  $f_{i}(\mathrm{C-H})=3.0\%$, implying that the bonds between the adsorbate and the surface atoms of NW-H are mostly covalent in nature.  
Hence, the presence of the surface-like states at the valence band edge in NW-F is not a spurious effect caused by the underestimation of the band gap within PBE, but is most likely  driven by the chemistry at the surface. The incomplete removal of the surface states from the valence band edge implies that fluorination might not produce the desired effects.  In what follows, we show this explicitly by creating $\mathrm{V_{Si}^{-1}}$ in the passivated NWs.
 
\vspace{-12pt}

\subsection{Silicon vacancy in passivated SiC NW}

\begin{figure*}[htb]
    \centering
    \includegraphics[width = 0.75\textwidth]{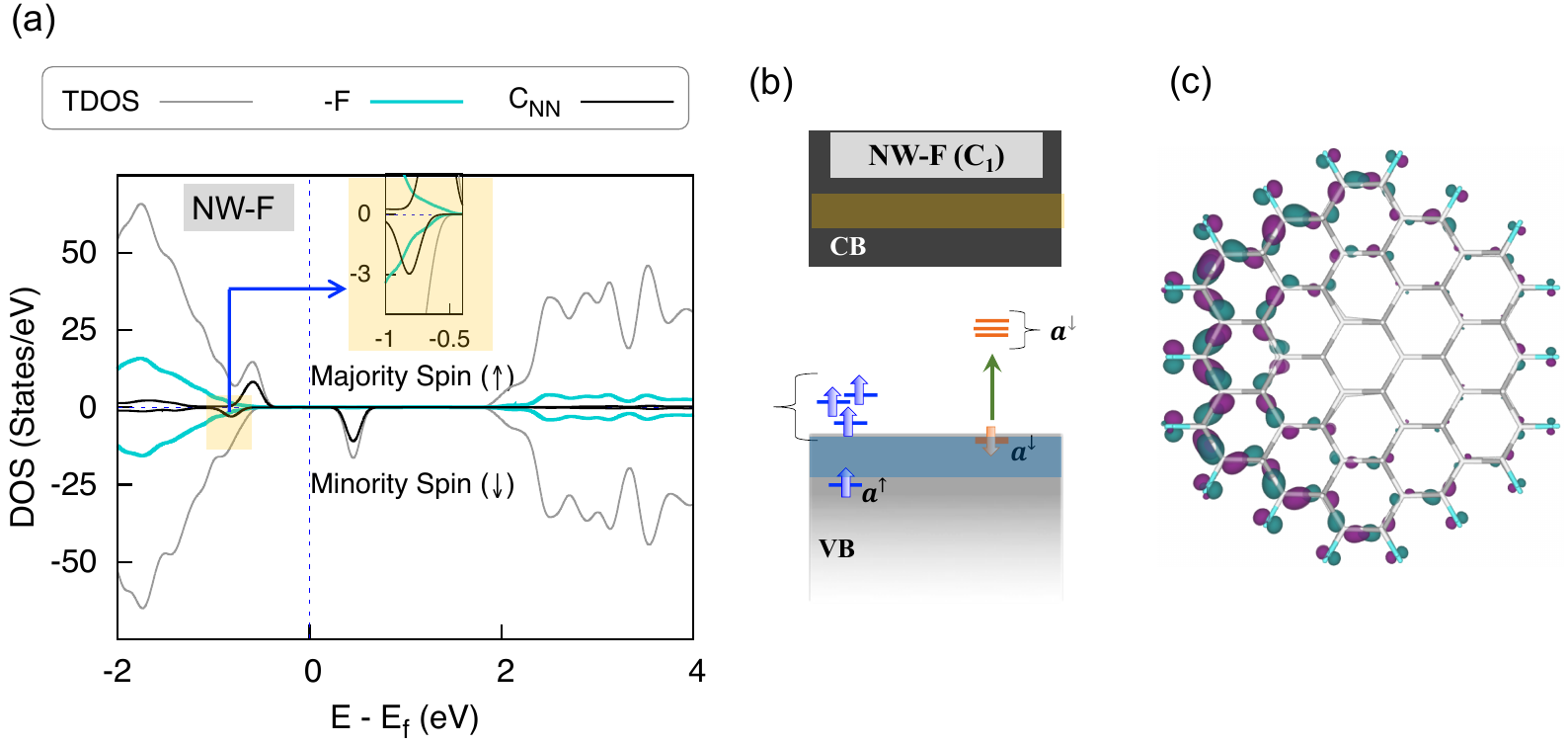}
        \vspace{-6pt}
    \caption{Fluorine-passivated 2H-SiC NW with a $\mathrm{V_{Si}^{-1}}$ defect. (a) Total DOS (TDOS) in gray and DOS projected onto the fluorine-atoms (in cyan) and the four carbon atoms that are the nearest neighbors ($\mathrm{C_{NN}}$) surrounding the defect. 
The filled optically-active state in the minority spin (spin-down) channel is resonant with the valence band as can be seen in the highlighted portion of the DOS, a zoomed-in version of which is shown in the inset. Above this filled minority spin state, one can see contributions from the fluorine atoms as well from other Si and C atoms in NW-F. (b) Schematic energy-level diagram (not to scale), highlighting how the electronic structure properties of a defect in the fluorinated NW differ from the other two passivation schemes. The filled optically-active state in the minority spin (spin-down) channel now lies below the valence band edge, which shows a mixed surface-bulk character.  This is depicted with a blue color band in VB. (c) The charge density plot of the state at the valence band edge, showing its hybrid surface-bulk character. 
}
    \label{fig:fig5}
\end{figure*}

\subsubsection{Ground state properties of $\mathrm{V_{Si}^{-1}}$ in passivated SiC NW}  
\vspace{-12pt}

A $\mathrm{V_{Si}^{-1}}$ defect was created in the interior of the NW to emulate near-surface defects. It is a spin-3/2 defect for all passivation schemes. 
The spin polarization energy, defined as the energy difference between the non-magnetic and magnetic structures, is found to be 713\,meV, 714\,meV and 694\,meV for hydrogen, mixed hydrogen/hydroxyl and fluorine passivations, respectively. These spin polarization energy values are much larger than the thermal energy at the room temperature, implying that the local moments induced by a $\mathrm{V_{Si}^{-1}}$ defect will survive well beyond room temperature in all three of the defective, passivated NWs.

Figures~\ref{fig:fig4} (a) and (b) are plots of the TDOS in gray and the projected DOS for defective NW-H and NW-H/OH. These plots clearly show a large spin-splitting between the majority and minority spin channels. The DOS contributions from the four nearest-neighboring carbon atoms, $\mathrm{C_{NN}}$, are plotted using black lines. The DOS projected onto  $\mathrm{C_{NN}}$ show that the defect states for $\mathrm{V_{Si}^{-1}}$ in nanowires -- NW-H and NW-H/OH -- are derived mostly from the $2s$ and $2p$ orbitals of $\mathrm{C_{NN}}$. The adsorbate-derived states show negligible contributions to the defect states. Additionally, the filled and empty optically-active defect states in the minority spin (spin-down) channel are now in the band gap, well separated from the valence and conduction bands. This is summarized in the schematic energy-level diagram (not to scale) shown in Fig.~\ref{fig:fig4} (c).  It highlights how the electronic structure properties of a defect in the passivated NWs  (jointly labelled as NW-X, with X = -H or -H/OH), differ from that for the bare NW. The schematic diagram shows that: (i) the filled and empty optically-active states in the minority spin (spin-down) channel are well-separated from the valence and conduction bands, and (ii) the dangling bond/surface states that previously occupied the band gap in the bare NW are now passivated and pushed deeper into valence and conduction bands (presented as colored bands within VB and CB).  Figure~\ref{fig:fig4} (d) shows the charge density plots of the optically-active empty and filled state for NW-H, showing the extent of spatial localization of the defect states around the defect site. The defect states for NW-H/OH with a $\mathrm{V_{Si}^{-1}}$ defect look similar and are not shown.  It should also be pointed out that since the defective structures have $C_1$ symmetries, all of the defect states are singlets and belong to the $a$-representation.

Figure~\ref{fig:fig5} (a) shows the TDOS (gray line) for the fluorine-passivated NW with a $\mathrm{V_{Si}^{-1}}$ defect. The DOS projected onto the fluorine-atoms (in cyan) and the four carbon atoms that are the nearest neighbors ($\mathrm{C_{NN}}$) surrounding the defect are also plotted, once again showing that it is the $2s$- and $2p$-orbitals of $\mathrm{C_{NN}}$ that contribute to the defect states introduced by $\mathrm{V_{Si}^{-1}}$. However, in the case of NW-F, the filled optically-active state in the minority spin (spin-down) channel is resonant with the valence band, as can be seen in the highlighted portion of the DOS, a zoomed-in version of which is shown in the inset in Fig.~\ref{fig:fig5} (a).  From the inset, one can see that the valence band edge itself has contributions from the fluorine atoms as well from other Si and C atoms in NW-F.  The results from the DOS plots are summarized in the schematic energy-level diagram (not to scale) shown in Fig.~\ref{fig:fig5} (b). It highlights how the electronic structure properties of a defect in NW-F differ from the other two passivation schemes. The filled optically-active state in the minority spin (spin-down) channel now lies below the valence band edge.  The DOS at the valence band edge have a mixed surface-bulk character, which is depicted with a blue color band in the VB. The mixed character can also be seen in the charge density plot of the state at the valence band edge. This filled state lies above the defect state involved in the photo-excitation. It may itself become involved in the photo-excitation of the $\mathrm{V_{Si}^{-1}}$ defect, and adversely affect the defect's optical properties. It may do so by  removing the hole left behind by the photoexcited electron from the defect site. These results show that fluorine passivation should be avoided. Hence, from here on we will only present our results for $\mathrm{V_{Si}^{-1}}$ in NW-H and NW-H/OH.
\begin{figure}[htb]
    \centering
    \includegraphics[width = 0.3\textwidth]{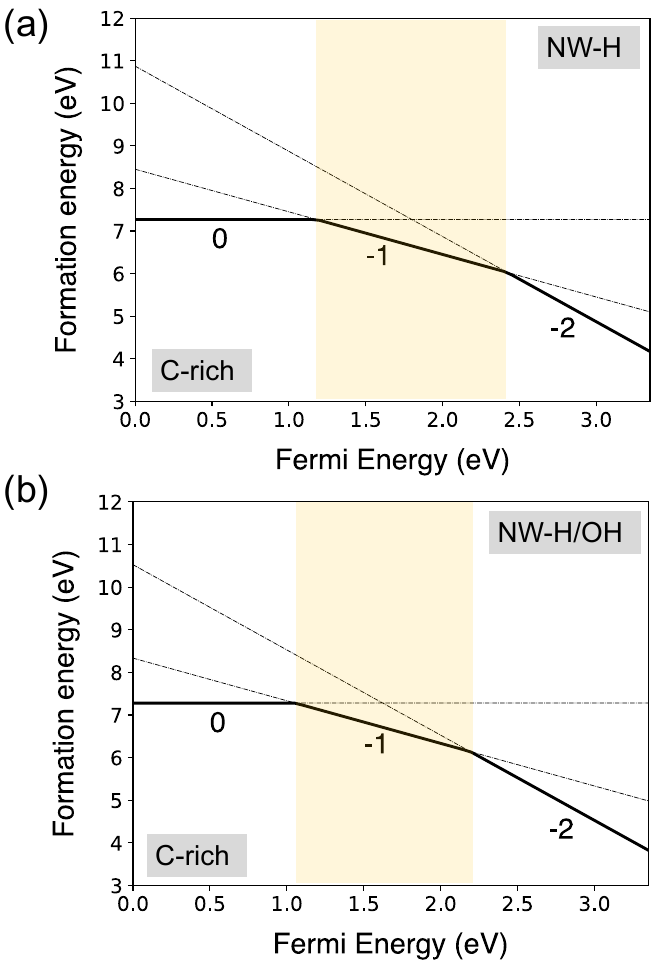}
    \vspace{-6pt}
    \caption{The calculated formation energies ($\Delta E_{form}$) for $\mathrm{V_{Si}}$ in different charged states as a function of the Fermi energy level (electronic chemical potential) in (a) NW-H and (b) NW-H/OH.  The Fermi energy level is given with respect to the valence band maximum.  The formation energies are calculated assuming carbon-rich conditions. The favorable doping conditions, corresponding to the values of electronic chemical potential at which the bright negatively charged state of the defect is stable, are highlighted. 
    }
    \label{fig:fig6}
\end{figure}

The DOS plots in Figs~\ref{fig:fig4} (a) and (b) clearly show that the surface-related states are removed from the respective band gaps in both NW-H and NW-H/OH by the covalent bonds formed by the adsorbates on the surfaces. This is one of the important requirements for an adsorbate to be considered optimal since the surface states are detrimental to the photostability of the near surface defects. However, it is also possible for a passivation scheme to be unfavorable if it reduces the doping-range (electronic chemical potential range) over which the negative charge state of the monovacancy is stable relative to its other charge states. Figures~\ref{fig:fig6} (a) and (b) show the calculated formation energies for $\mathrm{V_{Si}}$ in NW-H and NW-H/OH as a function of the Fermi energy level (i.e. the electronic chemical potential).  The Fermi energy level is given with respect to the valence band maximum.  In addition, since we are interested in growth conditions that result in formation of silicon monovacancies, the formation energies shown in Figs~\ref{fig:fig6} (a) and (b) are those obtained under carbon-rich (silicon-poor) conditions.  We have highlighted the favorable doping conditions in yellow, which correspond to the values of electronic chemical potential at which the bright negatively charged state of the defect is stable. The highlighted regions reveal that the bright, negatively-charged state of $\mathrm{V_{Si}}$ is stable over a considerable range of n-doping for both NW-H and NW-H/OH.  These results are almost identical to those obtained in an earlier work for a silicon monovacancy in bulk 4H-SiC~\cite{szasz2015,gordon2015_VSim1_HSE}, showing that the two passivation schemes being considered here are good choices for ensuring photostability of  $\mathrm{V_{Si}^{-1}}$.  

\vspace{-12pt}
\subsubsection{Excited state properties of $\mathrm{V_{Si}^{-1}}$ in passivated SiC NW}  
\vspace{-12pt}

To determine how the surface passivation affects the optical properties of the near-surface defects, we investigated  the excited state properties of $\mathrm{V_{Si}^{-1}}$ in NW-H and NW-H/OH. Here, we considered the lowest-energy, spin-preserving excitation between the highest filled and lowest empty defect states in the minority spin channel, as indicated by the vertical arrow in the schematic energy-level diagram in Fig.~\ref{fig:fig4} (c). Both states involved in this photoexcitation belong to the $a$-representation.  Our $\Delta$SCF results show that the ZPL of of $\mathrm{V_{Si}^{-1}}$ is 1.15\,eV for both NW-H and NW-H/OH.  This near-perfect restoration of the defect's ZPL to its value in the bulk 2H-SiC (1.19\,eV~\cite{Joshi2022}) can be attributed to the effective removal of the surface states from the energetic vicinity of the defect states. 


 In order to understand what contributes to the 40\,meV difference in the ZPL of $\mathrm{V_{Si}^{-1}}$ in NW-H and NW-H/OH as compared to the bulk value, we first consider the effect that the strain at the defect site has on the ZPL.  To estimate the change in ZPL due to strain, we use the strain-coupling parameter of 1130\,meV(strain$)^{-1}$ (in the basal direction), which was deduced experimentally~\cite{nanostr_VSi_expt_2020}.  In NW-H, the bond lengths along the basal plane (at the defect site) are around 0.18\% longer as compared to the ideal basal distance. A similar strain exists at the defect site in the case of NW-H/OH (0.21\%). Hence, for both passivations, strain results in a mere 2.03--2.04\,meV change in the ZPL.  Further investigation revealed that the change in the Stokes shift  is a more important contributor to the ZPL change in the passivated NW. The Stokes shift, which is obtained from the difference in the vertical excitation and the ZPL [i.e. $\mathrm{E_{B}-E_{C}}$ in Fig.~\ref{fig:fig2}(a)], is 104.9\,meV, 133.4\,meV and 128.7\,meV in the bulk, NW-H and NW-H/OH, respectively. This means that $\mathrm{V_{Si}^{-1}}$ undergoes larger structural changes upon photo-excitation in the passivated NW as compared to the bulk. The calculated values of Stokes shift are larger by about 28.5\,meV in NW-H and 23.7\,meV in NW-H/OH as compared to the Stokes shift in bulk.  Thus, the changes in Stokes shift can account for most of the calculated ZPL shift (about 40\,meV) between the  passivated NWs and the bulk.  A small contribution to the ZPL shift also comes from different placement of the defect states in the band gap due to symmetry breaking and changes in electronic structure properties that result from a particular surface chemistry.  As a result, the vertical excitation values, corresponding to the excitation $\mathrm{A \rightarrow B}$ in Fig.~\ref{fig:fig2}(a), differ by 11.2\,meV for NW-H and 19.0\,meV for NW-H/OH as compared to the bulk value for $\mathrm{E_{B}-E_{A}}$.  Hence, we find that the ZPL shifts in the passivated NWs are not only an order of magnitude smaller than that in the bare NW, but also the major contributor to this change is no longer the same. In the bare NW, hybridization of the defect states with the surface states accounted for most of the ZPL shift~\cite{Joshi2022}.  On the other hand, in the passivated NW, most of the shift can be accounted for by changes in Stokes shift, the placement of defect states within the band gap of a passivated NW and to a much lesser extent, by the strain.    
 


\section{Summary}

In this work, we have identified two chemical terminations for the surfaces of nanostructured SiC that effectively remove the surface states from the band gap of the nanostructure. These two passivation schemes involve either using uniform -H coverage or using a mixed -H/OH group coverage. Both coverages address the issue of charge-state instability of the near-surface $\mathrm{V_{Si}^{-1}}$ defect by effectively removing the surface states from the band gap of the nanostructure. The defect's high spin state (S=3/2) remains stable over a large range of electronic chemical potential in an n-doped SiC nanostructure.   We also show that surface passivation with fluorine  should be avoided since it is unable to completely eliminate surface-related states from the valence band edge.  In addition, we find that in NW-H and NW-H/OH, we have not only increased the photostability of $\mathrm{V_{Si}^{-1}}$, but there is also a near-perfect restoration of the frequency of its quantum emission. Lastly, although a small-diameter nanowire was used in our proof-of-principle calculations, the core results should remain valid for the spin-active defects that are placed advertently or inadvertently near the surfaces in larger nanostructures.



\vspace{12pt}
\begin{acknowledgments}

\noindent \textbf{Acknowledgements:} We acknowledge support by the National Science Foundation under NSF grant number DMR-1738076 and the STC Center for Integrated Quantum Materials under NSF Grant number DMR-1231319. This work used the Expanse and Bridges2 clusters at SDSC and PSC, respectively, through allocation PHY180014 from the Advanced Cyberinfrastructure Coordination Ecosystem: Services \& Support (ACCESS) program, which is supported by National Science Foundation grants No. 2138259, No. 2138286, No. 2138307, No. 2137603, and No. 2138296. 
Dr. Priyanka Manchanda is acknowledged for creating the passivated NW structures.  C.A.S.N. acknowledges Dr. Sharmila Shirodkar for useful discussions.


\end{acknowledgments}

\noindent \textbf{Author Contributions:} P.D. conceived and directed the study.  C.A.S.N., P.D. and T.J. performed ground state calculations.  P.D. and C.A.S.N. performed the $\Delta SCF$ calculations and wrote the manuscript.  All authors proof-read and reviewed the final revision of the manuscript.

\normalem  

\begin{thebibliography}{58}
\expandafter\ifx\csname natexlab\endcsname\relax\def\natexlab#1{#1}\fi
\expandafter\ifx\csname bibnamefont\endcsname\relax
  \def\bibnamefont#1{#1}\fi
\expandafter\ifx\csname bibfnamefont\endcsname\relax
  \def\bibfnamefont#1{#1}\fi
\expandafter\ifx\csname citenamefont\endcsname\relax
  \def\citenamefont#1{#1}\fi
\expandafter\ifx\csname url\endcsname\relax
  \def\url#1{\texttt{#1}}\fi
\expandafter\ifx\csname urlprefix\endcsname\relax\def\urlprefix{URL }\fi
\providecommand{\bibinfo}[2]{#2}
\providecommand{\eprint}[2][]{\url{#2}}

\bibitem[{\citenamefont{Falk et~al.}(2013)\citenamefont{Falk, Buckley,
  Calusine, Koehl, Dobrovitski, Politi, Zorman, Feng, and
  Awschalom}}]{falk2013polytype}
\bibinfo{author}{\bibfnamefont{A.~L.} \bibnamefont{Falk}},
  \bibinfo{author}{\bibfnamefont{B.~B.} \bibnamefont{Buckley}},
  \bibinfo{author}{\bibfnamefont{G.}~\bibnamefont{Calusine}},
  \bibinfo{author}{\bibfnamefont{W.~F.} \bibnamefont{Koehl}},
  \bibinfo{author}{\bibfnamefont{V.~V.} \bibnamefont{Dobrovitski}},
  \bibinfo{author}{\bibfnamefont{A.}~\bibnamefont{Politi}},
  \bibinfo{author}{\bibfnamefont{C.~A.} \bibnamefont{Zorman}},
  \bibinfo{author}{\bibfnamefont{P.~X.-L.} \bibnamefont{Feng}},
  \bibnamefont{and} \bibinfo{author}{\bibfnamefont{D.~D.}
  \bibnamefont{Awschalom}}, \bibinfo{journal}{Nature Communications}
  \textbf{\bibinfo{volume}{4}}, \bibinfo{pages}{1} (\bibinfo{year}{2013}),
  \urlprefix\url{https://doi.org/10.1038/ncomms2854}.

\bibitem[{\citenamefont{Widmann et~al.}(2015)\citenamefont{Widmann, Lee,
  Rendler, Son, Fedder, Paik, Yang, Zhao, Yang, Booker
  et~al.}}]{widmann2015coherentRT}
\bibinfo{author}{\bibfnamefont{M.}~\bibnamefont{Widmann}},
  \bibinfo{author}{\bibfnamefont{S.-Y.} \bibnamefont{Lee}},
  \bibinfo{author}{\bibfnamefont{T.}~\bibnamefont{Rendler}},
  \bibinfo{author}{\bibfnamefont{N.~T.} \bibnamefont{Son}},
  \bibinfo{author}{\bibfnamefont{H.}~\bibnamefont{Fedder}},
  \bibinfo{author}{\bibfnamefont{S.}~\bibnamefont{Paik}},
  \bibinfo{author}{\bibfnamefont{L.-P.} \bibnamefont{Yang}},
  \bibinfo{author}{\bibfnamefont{N.}~\bibnamefont{Zhao}},
  \bibinfo{author}{\bibfnamefont{S.}~\bibnamefont{Yang}},
  \bibinfo{author}{\bibfnamefont{I.}~\bibnamefont{Booker}},
  \bibnamefont{et~al.}, \bibinfo{journal}{Nature Materials}
  \textbf{\bibinfo{volume}{14}}, \bibinfo{pages}{164} (\bibinfo{year}{2015}),
  \urlprefix\url{https://doi.org/10.1038/nmat4145}.

\bibitem[{\citenamefont{Seo et~al.}(2016)\citenamefont{Seo, Falk, Klimov, Miao,
  Galli, and Awschalom}}]{Seo2016}
\bibinfo{author}{\bibfnamefont{H.}~\bibnamefont{Seo}},
  \bibinfo{author}{\bibfnamefont{A.~L.} \bibnamefont{Falk}},
  \bibinfo{author}{\bibfnamefont{P.~V.} \bibnamefont{Klimov}},
  \bibinfo{author}{\bibfnamefont{K.~C.} \bibnamefont{Miao}},
  \bibinfo{author}{\bibfnamefont{G.}~\bibnamefont{Galli}}, \bibnamefont{and}
  \bibinfo{author}{\bibfnamefont{D.~D.} \bibnamefont{Awschalom}},
  \bibinfo{journal}{Nature Communications} \textbf{\bibinfo{volume}{7}},
  \bibinfo{pages}{12935} (\bibinfo{year}{2016}),
  \urlprefix\url{https://doi.org/10.1038/ncomms12935}.

\bibitem[{\citenamefont{Nagy et~al.}(2019)\citenamefont{Nagy, Niethammer,
  Widmann, Chen, Udvarhelyi, Bonato, Hassan, Karhu, Ivanov, Son
  et~al.}}]{nagy2019highfidelity}
\bibinfo{author}{\bibfnamefont{R.}~\bibnamefont{Nagy}},
  \bibinfo{author}{\bibfnamefont{M.}~\bibnamefont{Niethammer}},
  \bibinfo{author}{\bibfnamefont{M.}~\bibnamefont{Widmann}},
  \bibinfo{author}{\bibfnamefont{Y.-C.} \bibnamefont{Chen}},
  \bibinfo{author}{\bibfnamefont{P.}~\bibnamefont{Udvarhelyi}},
  \bibinfo{author}{\bibfnamefont{C.}~\bibnamefont{Bonato}},
  \bibinfo{author}{\bibfnamefont{J.~U.} \bibnamefont{Hassan}},
  \bibinfo{author}{\bibfnamefont{R.}~\bibnamefont{Karhu}},
  \bibinfo{author}{\bibfnamefont{I.~G.} \bibnamefont{Ivanov}},
  \bibinfo{author}{\bibfnamefont{N.~T.} \bibnamefont{Son}},
  \bibnamefont{et~al.}, \bibinfo{journal}{Nature Communications}
  \textbf{\bibinfo{volume}{10}}, \bibinfo{pages}{1} (\bibinfo{year}{2019}),
  \urlprefix\url{https://doi.org/10.1038/s41467-019-09873-9}.

\bibitem[{\citenamefont{Christle et~al.}(2015)\citenamefont{Christle, Falk,
  Andrich, Klimov, Hassan, Son, Janzén, Ohshima, and
  Awschalom}}]{Christle2015}
\bibinfo{author}{\bibfnamefont{D.~J.} \bibnamefont{Christle}},
  \bibinfo{author}{\bibfnamefont{A.~L.} \bibnamefont{Falk}},
  \bibinfo{author}{\bibfnamefont{P.}~\bibnamefont{Andrich}},
  \bibinfo{author}{\bibfnamefont{P.~V.} \bibnamefont{Klimov}},
  \bibinfo{author}{\bibfnamefont{J.~U.} \bibnamefont{Hassan}},
  \bibinfo{author}{\bibfnamefont{N.~T.} \bibnamefont{Son}},
  \bibinfo{author}{\bibfnamefont{E.}~\bibnamefont{Janz\'{e}n}},
  \bibinfo{author}{\bibfnamefont{T.}~\bibnamefont{Ohshima}}, \bibnamefont{and}
  \bibinfo{author}{\bibfnamefont{D.~D.} \bibnamefont{Awschalom}},
  \bibinfo{journal}{Nature Materials} \textbf{\bibinfo{volume}{14}},
  \bibinfo{pages}{160} (\bibinfo{year}{2015}),
  \urlprefix\url{https://doi.org/10.1038/nmat4144}.

\bibitem[{\citenamefont{Carter et~al.}(2015)\citenamefont{Carter, Soykal, Dev,
  Economou, and Glaser}}]{Carter2015}
\bibinfo{author}{\bibfnamefont{S.~G.} \bibnamefont{Carter}},
  \bibinfo{author}{\bibfnamefont{O.~O.} \bibnamefont{Soykal}},
  \bibinfo{author}{\bibfnamefont{P.}~\bibnamefont{Dev}},
  \bibinfo{author}{\bibfnamefont{S.~E.} \bibnamefont{Economou}},
  \bibnamefont{and} \bibinfo{author}{\bibfnamefont{E.~R.}
  \bibnamefont{Glaser}}, \bibinfo{journal}{Physical Review B}
  \textbf{\bibinfo{volume}{92}}, \bibinfo{pages}{161202}
  (\bibinfo{year}{2015}),
  \urlprefix\url{https://link.aps.org/doi/10.1103/PhysRevB.92.161202}.

\bibitem[{\citenamefont{Udvarhelyi et~al.}(2019)\citenamefont{Udvarhelyi, Nagy,
  Kaiser, Lee, Wrachtrup, and Gali}}]{Udvarhelyi2019}
\bibinfo{author}{\bibfnamefont{P.}~\bibnamefont{Udvarhelyi}},
  \bibinfo{author}{\bibfnamefont{R.}~\bibnamefont{Nagy}},
  \bibinfo{author}{\bibfnamefont{F.}~\bibnamefont{Kaiser}},
  \bibinfo{author}{\bibfnamefont{S.-Y.} \bibnamefont{Lee}},
  \bibinfo{author}{\bibfnamefont{J.}~\bibnamefont{Wrachtrup}},
  \bibnamefont{and} \bibinfo{author}{\bibfnamefont{A.}~\bibnamefont{Gali}},
  \bibinfo{journal}{Physical Review Applied} \textbf{\bibinfo{volume}{11}},
  \bibinfo{pages}{044022} (\bibinfo{year}{2019}),
  \urlprefix\url{https://link.aps.org/doi/10.1103/PhysRevApplied.11.044022}.

\bibitem[{\citenamefont{Babin et~al.}(2022)\citenamefont{Babin, St{\"o}hr,
  Morioka, Linkewitz, Steidl, W{\"o}rnle, Liu, Hesselmeier, Vorobyov, Denisenko
  et~al.}}]{babin2021nanofabricated}
\bibinfo{author}{\bibfnamefont{C.}~\bibnamefont{Babin}},
  \bibinfo{author}{\bibfnamefont{R.}~\bibnamefont{St{\"o}hr}},
  \bibinfo{author}{\bibfnamefont{N.}~\bibnamefont{Morioka}},
  \bibinfo{author}{\bibfnamefont{T.}~\bibnamefont{Linkewitz}},
  \bibinfo{author}{\bibfnamefont{T.}~\bibnamefont{Steidl}},
  \bibinfo{author}{\bibfnamefont{R.}~\bibnamefont{W{\"o}rnle}},
  \bibinfo{author}{\bibfnamefont{D.}~\bibnamefont{Liu}},
  \bibinfo{author}{\bibfnamefont{E.}~\bibnamefont{Hesselmeier}},
  \bibinfo{author}{\bibfnamefont{V.}~\bibnamefont{Vorobyov}},
  \bibinfo{author}{\bibfnamefont{A.}~\bibnamefont{Denisenko}},
  \bibnamefont{et~al.}, \bibinfo{journal}{Nature Materials}
  \textbf{\bibinfo{volume}{21}}, \bibinfo{pages}{67} (\bibinfo{year}{2022}),
  ISSN \bibinfo{issn}{1476-4660},
  \urlprefix\url{https://doi.org/10.1038/s41563-021-01148-3}.

\bibitem[{\citenamefont{Fuchs et~al.}(2015)\citenamefont{Fuchs, Stender,
  Trupke, Simin, Pflaum, Dyakonov, and Astakhov}}]{Fuchs2015}
\bibinfo{author}{\bibfnamefont{F.}~\bibnamefont{Fuchs}},
  \bibinfo{author}{\bibfnamefont{B.}~\bibnamefont{Stender}},
  \bibinfo{author}{\bibfnamefont{M.}~\bibnamefont{Trupke}},
  \bibinfo{author}{\bibfnamefont{D.}~\bibnamefont{Simin}},
  \bibinfo{author}{\bibfnamefont{J.}~\bibnamefont{Pflaum}},
  \bibinfo{author}{\bibfnamefont{V.}~\bibnamefont{Dyakonov}}, \bibnamefont{and}
  \bibinfo{author}{\bibfnamefont{G.~V.} \bibnamefont{Astakhov}},
  \bibinfo{journal}{Nature Communications} \textbf{\bibinfo{volume}{6}},
  \bibinfo{pages}{7578} (\bibinfo{year}{2015}),
  \urlprefix\url{https://doi.org/10.1038/ncomms8578}.

\bibitem[{\citenamefont{Balasubramanian
  et~al.}(2008)\citenamefont{Balasubramanian, Chan, Kolesov, Al-Hmoud, Tisler,
  Shin, Kim, Wojcik, Hemmer, Krueger et~al.}}]{balasubramanian2008}
\bibinfo{author}{\bibfnamefont{G.}~\bibnamefont{Balasubramanian}},
  \bibinfo{author}{\bibfnamefont{I.~Y.} \bibnamefont{Chan}},
  \bibinfo{author}{\bibfnamefont{R.}~\bibnamefont{Kolesov}},
  \bibinfo{author}{\bibfnamefont{M.}~\bibnamefont{Al-Hmoud}},
  \bibinfo{author}{\bibfnamefont{J.}~\bibnamefont{Tisler}},
  \bibinfo{author}{\bibfnamefont{C.}~\bibnamefont{Shin}},
  \bibinfo{author}{\bibfnamefont{C.}~\bibnamefont{Kim}},
  \bibinfo{author}{\bibfnamefont{A.}~\bibnamefont{Wojcik}},
  \bibinfo{author}{\bibfnamefont{P.~R.} \bibnamefont{Hemmer}},
  \bibinfo{author}{\bibfnamefont{A.}~\bibnamefont{Krueger}},
  \bibnamefont{et~al.}, \bibinfo{journal}{Nature}
  \textbf{\bibinfo{volume}{455}}, \bibinfo{pages}{648} (\bibinfo{year}{2008}),
  ISSN \bibinfo{issn}{1476-4687},
  \urlprefix\url{https://doi.org/10.1038/nature07278}.

\bibitem[{\citenamefont{Gali et~al.}(2009)\citenamefont{Gali, Janz\'en, De\'ak,
  Kresse, and Kaxiras}}]{KaxirasNV}
\bibinfo{author}{\bibfnamefont{A.}~\bibnamefont{Gali}},
  \bibinfo{author}{\bibfnamefont{E.}~\bibnamefont{Janz\'en}},
  \bibinfo{author}{\bibfnamefont{P.}~\bibnamefont{De\'ak}},
  \bibinfo{author}{\bibfnamefont{G.}~\bibnamefont{Kresse}}, \bibnamefont{and}
  \bibinfo{author}{\bibfnamefont{E.}~\bibnamefont{Kaxiras}},
  \bibinfo{journal}{Phys. Rev. Lett.} \textbf{\bibinfo{volume}{103}},
  \bibinfo{pages}{186404} (\bibinfo{year}{2009}),
  \urlprefix\url{https://link.aps.org/doi/10.1103/PhysRevLett.103.186404}.

\bibitem[{\citenamefont{Stanwix et~al.}(2010)\citenamefont{Stanwix, Pham, Maze,
  Le~Sage, Yeung, Cappellaro, Hemmer, Yacoby, Lukin, and
  Walsworth}}]{stanwix2010}
\bibinfo{author}{\bibfnamefont{P.~L.} \bibnamefont{Stanwix}},
  \bibinfo{author}{\bibfnamefont{L.~M.} \bibnamefont{Pham}},
  \bibinfo{author}{\bibfnamefont{J.~R.} \bibnamefont{Maze}},
  \bibinfo{author}{\bibfnamefont{D.}~\bibnamefont{Le~Sage}},
  \bibinfo{author}{\bibfnamefont{T.~K.} \bibnamefont{Yeung}},
  \bibinfo{author}{\bibfnamefont{P.}~\bibnamefont{Cappellaro}},
  \bibinfo{author}{\bibfnamefont{P.~R.} \bibnamefont{Hemmer}},
  \bibinfo{author}{\bibfnamefont{A.}~\bibnamefont{Yacoby}},
  \bibinfo{author}{\bibfnamefont{M.~D.} \bibnamefont{Lukin}}, \bibnamefont{and}
  \bibinfo{author}{\bibfnamefont{R.~L.} \bibnamefont{Walsworth}},
  \bibinfo{journal}{Phys. Rev. B} \textbf{\bibinfo{volume}{82}},
  \bibinfo{pages}{201201} (\bibinfo{year}{2010}),
  \urlprefix\url{https://link.aps.org/doi/10.1103/PhysRevB.82.201201}.

\bibitem[{\citenamefont{Abtew et~al.}(2011)\citenamefont{Abtew, Sun, Shih, Dev,
  Zhang, and Zhang}}]{tesfaye2011}
\bibinfo{author}{\bibfnamefont{T.~A.} \bibnamefont{Abtew}},
  \bibinfo{author}{\bibfnamefont{Y.~Y.} \bibnamefont{Sun}},
  \bibinfo{author}{\bibfnamefont{B.-C.} \bibnamefont{Shih}},
  \bibinfo{author}{\bibfnamefont{P.}~\bibnamefont{Dev}},
  \bibinfo{author}{\bibfnamefont{S.~B.} \bibnamefont{Zhang}}, \bibnamefont{and}
  \bibinfo{author}{\bibfnamefont{P.}~\bibnamefont{Zhang}},
  \bibinfo{journal}{Phys. Rev. Lett.} \textbf{\bibinfo{volume}{107}},
  \bibinfo{pages}{146403} (\bibinfo{year}{2011}),
  \urlprefix\url{https://link.aps.org/doi/10.1103/PhysRevLett.107.146403}.

\bibitem[{\citenamefont{Doherty et~al.}(2013)\citenamefont{Doherty, Manson,
  Delaney, Jelezko, Wrachtrup, and Hollenberg}}]{doherty2013nitrogen}
\bibinfo{author}{\bibfnamefont{M.~W.} \bibnamefont{Doherty}},
  \bibinfo{author}{\bibfnamefont{N.~B.} \bibnamefont{Manson}},
  \bibinfo{author}{\bibfnamefont{P.}~\bibnamefont{Delaney}},
  \bibinfo{author}{\bibfnamefont{F.}~\bibnamefont{Jelezko}},
  \bibinfo{author}{\bibfnamefont{J.}~\bibnamefont{Wrachtrup}},
  \bibnamefont{and} \bibinfo{author}{\bibfnamefont{L.~C.}
  \bibnamefont{Hollenberg}}, \bibinfo{journal}{Physics Reports}
  \textbf{\bibinfo{volume}{528}}, \bibinfo{pages}{1} (\bibinfo{year}{2013}),
  \urlprefix\url{https://www.sciencedirect.com/science/article/pii/S0370157313000562}.

\bibitem[{\citenamefont{Bhandari et~al.}(2021)\citenamefont{Bhandari, Wysocki,
  Economou, Dev, and Park}}]{bhandari2021}
\bibinfo{author}{\bibfnamefont{C.}~\bibnamefont{Bhandari}},
  \bibinfo{author}{\bibfnamefont{A.~L.} \bibnamefont{Wysocki}},
  \bibinfo{author}{\bibfnamefont{S.~E.} \bibnamefont{Economou}},
  \bibinfo{author}{\bibfnamefont{P.}~\bibnamefont{Dev}}, \bibnamefont{and}
  \bibinfo{author}{\bibfnamefont{K.}~\bibnamefont{Park}},
  \bibinfo{journal}{Phys. Rev. B} \textbf{\bibinfo{volume}{103}},
  \bibinfo{pages}{014115} (\bibinfo{year}{2021}),
  \urlprefix\url{https://link.aps.org/doi/10.1103/PhysRevB.103.014115}.

\bibitem[{\citenamefont{Gadalla et~al.}(2021)\citenamefont{Gadalla, Greenspon,
  Defo, Zhang, and Hu}}]{gadalla2021enhanced}
\bibinfo{author}{\bibfnamefont{M.~N.} \bibnamefont{Gadalla}},
  \bibinfo{author}{\bibfnamefont{A.~S.} \bibnamefont{Greenspon}},
  \bibinfo{author}{\bibfnamefont{R.~K.} \bibnamefont{Defo}},
  \bibinfo{author}{\bibfnamefont{X.}~\bibnamefont{Zhang}}, \bibnamefont{and}
  \bibinfo{author}{\bibfnamefont{E.~L.} \bibnamefont{Hu}},
  \bibinfo{journal}{Proceedings of the National Academy of Sciences}
  \textbf{\bibinfo{volume}{118}} (\bibinfo{year}{2021}),
  \urlprefix\url{https://doi.org/10.1073/pnas.2021768118}.

\bibitem[{\citenamefont{Majety et~al.}(2021)\citenamefont{Majety, Norman, Li,
  Bell, Saha, and Radulaski}}]{majety2021quantum}
\bibinfo{author}{\bibfnamefont{S.}~\bibnamefont{Majety}},
  \bibinfo{author}{\bibfnamefont{V.~A.} \bibnamefont{Norman}},
  \bibinfo{author}{\bibfnamefont{L.}~\bibnamefont{Li}},
  \bibinfo{author}{\bibfnamefont{M.}~\bibnamefont{Bell}},
  \bibinfo{author}{\bibfnamefont{P.}~\bibnamefont{Saha}}, \bibnamefont{and}
  \bibinfo{author}{\bibfnamefont{M.}~\bibnamefont{Radulaski}},
  \bibinfo{journal}{Journal of Physics: Photonics}
  \textbf{\bibinfo{volume}{3}}, \bibinfo{pages}{034008} (\bibinfo{year}{2021}),
  \urlprefix\url{https://doi.org/10.1088/2515-7647/abfdca}.

\bibitem[{\citenamefont{Crook et~al.}(2020)\citenamefont{Crook, Anderson, Miao,
  Bourassa, Lee, Bayliss, Bracher, Zhang, Abe, Ohshima
  et~al.}}]{crookEHu2020purcell}
\bibinfo{author}{\bibfnamefont{A.~L.} \bibnamefont{Crook}},
  \bibinfo{author}{\bibfnamefont{C.~P.} \bibnamefont{Anderson}},
  \bibinfo{author}{\bibfnamefont{K.~C.} \bibnamefont{Miao}},
  \bibinfo{author}{\bibfnamefont{A.}~\bibnamefont{Bourassa}},
  \bibinfo{author}{\bibfnamefont{H.}~\bibnamefont{Lee}},
  \bibinfo{author}{\bibfnamefont{S.~L.} \bibnamefont{Bayliss}},
  \bibinfo{author}{\bibfnamefont{D.~O.} \bibnamefont{Bracher}},
  \bibinfo{author}{\bibfnamefont{X.}~\bibnamefont{Zhang}},
  \bibinfo{author}{\bibfnamefont{H.}~\bibnamefont{Abe}},
  \bibinfo{author}{\bibfnamefont{T.}~\bibnamefont{Ohshima}},
  \bibnamefont{et~al.}, \bibinfo{journal}{Nano Letters}
  \textbf{\bibinfo{volume}{20}}, \bibinfo{pages}{3427} (\bibinfo{year}{2020}),
  \urlprefix\url{https://doi.org/10.1021/acs.nanolett.0c00339}.

\bibitem[{\citenamefont{Bracher et~al.}(2017)\citenamefont{Bracher, Zhang, and
  Hu}}]{bracherEHu2017selective}
\bibinfo{author}{\bibfnamefont{D.~O.} \bibnamefont{Bracher}},
  \bibinfo{author}{\bibfnamefont{X.}~\bibnamefont{Zhang}}, \bibnamefont{and}
  \bibinfo{author}{\bibfnamefont{E.~L.} \bibnamefont{Hu}},
  \bibinfo{journal}{Proceedings of the National Academy of Sciences}
  \textbf{\bibinfo{volume}{114}}, \bibinfo{pages}{4060} (\bibinfo{year}{2017}),
  \urlprefix\url{https://doi.org/10.1073/pnas.1704219114}.

\bibitem[{\citenamefont{Nagy et~al.}(2018)\citenamefont{Nagy, Widmann,
  Niethammer, Dasari, Gerhardt, Soykal, Radulaski, Ohshima, Vu{\v{c}}kovi{\'c},
  Son et~al.}}]{nagy2018quantumdichroic}
\bibinfo{author}{\bibfnamefont{R.}~\bibnamefont{Nagy}},
  \bibinfo{author}{\bibfnamefont{M.}~\bibnamefont{Widmann}},
  \bibinfo{author}{\bibfnamefont{M.}~\bibnamefont{Niethammer}},
  \bibinfo{author}{\bibfnamefont{D.~B.} \bibnamefont{Dasari}},
  \bibinfo{author}{\bibfnamefont{I.}~\bibnamefont{Gerhardt}},
  \bibinfo{author}{\bibfnamefont{{\"O}.~O.} \bibnamefont{Soykal}},
  \bibinfo{author}{\bibfnamefont{M.}~\bibnamefont{Radulaski}},
  \bibinfo{author}{\bibfnamefont{T.}~\bibnamefont{Ohshima}},
  \bibinfo{author}{\bibfnamefont{J.}~\bibnamefont{Vu{\v{c}}kovi{\'c}}},
  \bibinfo{author}{\bibfnamefont{N.~T.} \bibnamefont{Son}},
  \bibnamefont{et~al.}, \bibinfo{journal}{Physical Review Applied}
  \textbf{\bibinfo{volume}{9}}, \bibinfo{pages}{034022} (\bibinfo{year}{2018}),
  \urlprefix\url{https://doi.org/10.1103/PhysRevApplied.9.034022}.

\bibitem[{\citenamefont{Lukin et~al.}(2020)\citenamefont{Lukin, Dory, Guidry,
  Yang, Mishra, Trivedi, Radulaski, Sun, Vercruysse, Ahn
  et~al.}}]{lukinRadulaski20204h}
\bibinfo{author}{\bibfnamefont{D.~M.} \bibnamefont{Lukin}},
  \bibinfo{author}{\bibfnamefont{C.}~\bibnamefont{Dory}},
  \bibinfo{author}{\bibfnamefont{M.~A.} \bibnamefont{Guidry}},
  \bibinfo{author}{\bibfnamefont{K.~Y.} \bibnamefont{Yang}},
  \bibinfo{author}{\bibfnamefont{S.~D.} \bibnamefont{Mishra}},
  \bibinfo{author}{\bibfnamefont{R.}~\bibnamefont{Trivedi}},
  \bibinfo{author}{\bibfnamefont{M.}~\bibnamefont{Radulaski}},
  \bibinfo{author}{\bibfnamefont{S.}~\bibnamefont{Sun}},
  \bibinfo{author}{\bibfnamefont{D.}~\bibnamefont{Vercruysse}},
  \bibinfo{author}{\bibfnamefont{G.~H.} \bibnamefont{Ahn}},
  \bibnamefont{et~al.}, \bibinfo{journal}{Nature Photonics}
  \textbf{\bibinfo{volume}{14}}, \bibinfo{pages}{330} (\bibinfo{year}{2020}),
  \urlprefix\url{https://doi.org/10.1038/s41566-019-0556-6}.

\bibitem[{\citenamefont{Radulaski et~al.}(2017)\citenamefont{Radulaski,
  Widmann, Niethammer, Zhang, Lee, Rendler, Lagoudakis, Son, Janzen, Ohshima
  et~al.}}]{radulaski2017scalable}
\bibinfo{author}{\bibfnamefont{M.}~\bibnamefont{Radulaski}},
  \bibinfo{author}{\bibfnamefont{M.}~\bibnamefont{Widmann}},
  \bibinfo{author}{\bibfnamefont{M.}~\bibnamefont{Niethammer}},
  \bibinfo{author}{\bibfnamefont{J.~L.} \bibnamefont{Zhang}},
  \bibinfo{author}{\bibfnamefont{S.-Y.} \bibnamefont{Lee}},
  \bibinfo{author}{\bibfnamefont{T.}~\bibnamefont{Rendler}},
  \bibinfo{author}{\bibfnamefont{K.~G.} \bibnamefont{Lagoudakis}},
  \bibinfo{author}{\bibfnamefont{N.~T.} \bibnamefont{Son}},
  \bibinfo{author}{\bibfnamefont{E.}~\bibnamefont{Janzen}},
  \bibinfo{author}{\bibfnamefont{T.}~\bibnamefont{Ohshima}},
  \bibnamefont{et~al.}, \bibinfo{journal}{Nano Letters}
  \textbf{\bibinfo{volume}{17}}, \bibinfo{pages}{1782} (\bibinfo{year}{2017}),
  \urlprefix\url{https://doi.org/10.1021/acs.nanolett.6b05102}.

\bibitem[{\citenamefont{Yuan et~al.}(2020)\citenamefont{Yuan, Fitzpatrick,
  Rodgers, Sangtawesin, Srinivasan, and de~Leon}}]{YuanDeLeon2020Charge}
\bibinfo{author}{\bibfnamefont{Z.}~\bibnamefont{Yuan}},
  \bibinfo{author}{\bibfnamefont{M.}~\bibnamefont{Fitzpatrick}},
  \bibinfo{author}{\bibfnamefont{L.~V.~H.} \bibnamefont{Rodgers}},
  \bibinfo{author}{\bibfnamefont{S.}~\bibnamefont{Sangtawesin}},
  \bibinfo{author}{\bibfnamefont{S.}~\bibnamefont{Srinivasan}},
  \bibnamefont{and} \bibinfo{author}{\bibfnamefont{N.~P.}
  \bibnamefont{de~Leon}}, \bibinfo{journal}{Physical Review Research}
  \textbf{\bibinfo{volume}{2}}, \bibinfo{pages}{033263} (\bibinfo{year}{2020}),
  \urlprefix\url{https://link.aps.org/doi/10.1103/PhysRevResearch.2.033263}.

\bibitem[{\citenamefont{Sangtawesin et~al.}(2019)\citenamefont{Sangtawesin,
  Dwyer, Srinivasan, Allred, Rodgers, De~Greve, Stacey, Dontschuk, O?Donnell,
  Hu et~al.}}]{sangtawesinDeLeon2019origins}
\bibinfo{author}{\bibfnamefont{S.}~\bibnamefont{Sangtawesin}},
  \bibinfo{author}{\bibfnamefont{B.~L.} \bibnamefont{Dwyer}},
  \bibinfo{author}{\bibfnamefont{S.}~\bibnamefont{Srinivasan}},
  \bibinfo{author}{\bibfnamefont{J.~J.} \bibnamefont{Allred}},
  \bibinfo{author}{\bibfnamefont{L.~V.} \bibnamefont{Rodgers}},
  \bibinfo{author}{\bibfnamefont{K.}~\bibnamefont{De~Greve}},
  \bibinfo{author}{\bibfnamefont{A.}~\bibnamefont{Stacey}},
  \bibinfo{author}{\bibfnamefont{N.}~\bibnamefont{Dontschuk}},
  \bibinfo{author}{\bibfnamefont{K.~M.} \bibnamefont{O?Donnell}},
  \bibinfo{author}{\bibfnamefont{D.}~\bibnamefont{Hu}}, \bibnamefont{et~al.},
  \bibinfo{journal}{Physical Review X} \textbf{\bibinfo{volume}{9}},
  \bibinfo{pages}{031052} (\bibinfo{year}{2019}),
  \urlprefix\url{https://doi.org/10.1103/PhysRevX.9.031052}.

\bibitem[{\citenamefont{V{\'a}squez et~al.}(2020)\citenamefont{V{\'a}squez,
  Bathen, Galeckas, Bazioti, Johansen, Maestre, Cremades, Prytz, Moe, Kuznetsov
  et~al.}}]{nanostr_VSi_expt_2020}
\bibinfo{author}{\bibfnamefont{G.~C.} \bibnamefont{V{\'a}squez}},
  \bibinfo{author}{\bibfnamefont{M.~E.} \bibnamefont{Bathen}},
  \bibinfo{author}{\bibfnamefont{A.}~\bibnamefont{Galeckas}},
  \bibinfo{author}{\bibfnamefont{C.}~\bibnamefont{Bazioti}},
  \bibinfo{author}{\bibfnamefont{K.~M.} \bibnamefont{Johansen}},
  \bibinfo{author}{\bibfnamefont{D.}~\bibnamefont{Maestre}},
  \bibinfo{author}{\bibfnamefont{A.}~\bibnamefont{Cremades}},
  \bibinfo{author}{\bibfnamefont{{\O}.}~\bibnamefont{Prytz}},
  \bibinfo{author}{\bibfnamefont{A.~M.} \bibnamefont{Moe}},
  \bibinfo{author}{\bibfnamefont{A.~Y.} \bibnamefont{Kuznetsov}},
  \bibnamefont{et~al.}, \bibinfo{journal}{Nano Letters}
  \textbf{\bibinfo{volume}{20}}, \bibinfo{pages}{8689} (\bibinfo{year}{2020}),
  \urlprefix\url{https://doi.org/10.1021/acs.nanolett.0c03472}.

\bibitem[{\citenamefont{Neethirajan et~al.}(2023)\citenamefont{Neethirajan,
  Hache, Paone, Pinto, Denisenko, Stöhr, Udvarhelyi, Pershin, Gali, Wrachtrup
  et~al.}}]{neethirajan2023}
\bibinfo{author}{\bibfnamefont{J.~N.} \bibnamefont{Neethirajan}},
  \bibinfo{author}{\bibfnamefont{T.}~\bibnamefont{Hache}},
  \bibinfo{author}{\bibfnamefont{D.}~\bibnamefont{Paone}},
  \bibinfo{author}{\bibfnamefont{D.}~\bibnamefont{Pinto}},
  \bibinfo{author}{\bibfnamefont{A.}~\bibnamefont{Denisenko}},
  \bibinfo{author}{\bibfnamefont{R.}~\bibnamefont{St\"{o}hr}},
  \bibinfo{author}{\bibfnamefont{P.}~\bibnamefont{Udvarhelyi}},
  \bibinfo{author}{\bibfnamefont{A.}~\bibnamefont{Pershin}},
  \bibinfo{author}{\bibfnamefont{A.}~\bibnamefont{Gali}},
  \bibinfo{author}{\bibfnamefont{J.}~\bibnamefont{Wrachtrup}},
  \bibnamefont{et~al.}, \bibinfo{journal}{Nano Letters}
  \textbf{\bibinfo{volume}{23}}, \bibinfo{pages}{2563} (\bibinfo{year}{2023}),
  \urlprefix\url{https://doi.org/10.1021/acs.nanolett.2c04733}.

\bibitem[{\citenamefont{Kaviani et~al.}(2014)\citenamefont{Kaviani, Deák,
  Aradi, Frauenheim, Chou, and Gali}}]{Kaviani2014}
\bibinfo{author}{\bibfnamefont{M.}~\bibnamefont{Kaviani}},
  \bibinfo{author}{\bibfnamefont{P.}~\bibnamefont{De\'{a}k}},
  \bibinfo{author}{\bibfnamefont{B.}~\bibnamefont{Aradi}},
  \bibinfo{author}{\bibfnamefont{T.}~\bibnamefont{Frauenheim}},
  \bibinfo{author}{\bibfnamefont{J.-P.} \bibnamefont{Chou}}, \bibnamefont{and}
  \bibinfo{author}{\bibfnamefont{A.}~\bibnamefont{Gali}},
  \bibinfo{journal}{Nano Letters} \textbf{\bibinfo{volume}{14}},
  \bibinfo{pages}{4772} (\bibinfo{year}{2014}),
  \urlprefix\url{https://doi.org/10.1021/nl501927y}.

\bibitem[{\citenamefont{Li et~al.}(2019)\citenamefont{Li, Chou, Wei, Sun, Hu,
  and Gali}}]{LiGali2019}
\bibinfo{author}{\bibfnamefont{S.}~\bibnamefont{Li}},
  \bibinfo{author}{\bibfnamefont{J.-P.} \bibnamefont{Chou}},
  \bibinfo{author}{\bibfnamefont{J.}~\bibnamefont{Wei}},
  \bibinfo{author}{\bibfnamefont{M.}~\bibnamefont{Sun}},
  \bibinfo{author}{\bibfnamefont{A.}~\bibnamefont{Hu}}, \bibnamefont{and}
  \bibinfo{author}{\bibfnamefont{A.}~\bibnamefont{Gali}},
  \bibinfo{journal}{Carbon} \textbf{\bibinfo{volume}{145}},
  \bibinfo{pages}{273} (\bibinfo{year}{2019}), ISSN \bibinfo{issn}{0008-6223},
  \urlprefix\url{https://www.sciencedirect.com/science/article/pii/S0008622319300168}.

\bibitem[{\citenamefont{L\"{o}fgren et~al.}(2019)\citenamefont{L\"{o}fgren,
  Pawar, Öberg, and Larsson}}]{Lofgren_2019}
\bibinfo{author}{\bibfnamefont{R.}~\bibnamefont{L\"{o}fgren}},
  \bibinfo{author}{\bibfnamefont{R.}~\bibnamefont{Pawar}},
  \bibinfo{author}{\bibfnamefont{S.}~\bibnamefont{\"{O}berg}}, \bibnamefont{and}
  \bibinfo{author}{\bibfnamefont{J.~A.} \bibnamefont{Larsson}},
  \bibinfo{journal}{New Journal of Physics} \textbf{\bibinfo{volume}{21}},
  \bibinfo{pages}{053037} (\bibinfo{year}{2019}),
  \urlprefix\url{https://dx.doi.org/10.1088/1367-2630/ab1ec5}.

\bibitem[{\citenamefont{Joshi and Dev}(2022)}]{Joshi2022}
\bibinfo{author}{\bibfnamefont{T.}~\bibnamefont{Joshi}} \bibnamefont{and}
  \bibinfo{author}{\bibfnamefont{P.}~\bibnamefont{Dev}}, \bibinfo{journal}{PRX
  Quantum} \textbf{\bibinfo{volume}{3}}, \bibinfo{pages}{020325}
  (\bibinfo{year}{2022}),
  \urlprefix\url{https://link.aps.org/doi/10.1103/PRXQuantum.3.020325}.

\bibitem[{\citenamefont{Giannozzi et~al.}(2009)\citenamefont{Giannozzi, Baroni,
  Bonini, Calandra, Car, Cavazzoni, Ceresoli, Chiarotti, Cococcioni, Dabo
  et~al.}}]{QE-2009}
\bibinfo{author}{\bibfnamefont{P.}~\bibnamefont{Giannozzi}},
  \bibinfo{author}{\bibfnamefont{S.}~\bibnamefont{Baroni}},
  \bibinfo{author}{\bibfnamefont{N.}~\bibnamefont{Bonini}},
  \bibinfo{author}{\bibfnamefont{M.}~\bibnamefont{Calandra}},
  \bibinfo{author}{\bibfnamefont{R.}~\bibnamefont{Car}},
  \bibinfo{author}{\bibfnamefont{C.}~\bibnamefont{Cavazzoni}},
  \bibinfo{author}{\bibfnamefont{D.}~\bibnamefont{Ceresoli}},
  \bibinfo{author}{\bibfnamefont{G.~L.} \bibnamefont{Chiarotti}},
  \bibinfo{author}{\bibfnamefont{M.}~\bibnamefont{Cococcioni}},
  \bibinfo{author}{\bibfnamefont{I.}~\bibnamefont{Dabo}}, \bibnamefont{et~al.},
  \bibinfo{journal}{Journal of Physics: Condensed Matter}
  \textbf{\bibinfo{volume}{21}}, \bibinfo{pages}{395502 (19pp)}
  (\bibinfo{year}{2009}), \urlprefix\url{http://www.quantum-espresso.org}.

\bibitem[{\citenamefont{Giannozzi et~al.}(2017)\citenamefont{Giannozzi,
  Andreussi, Brumme, Bunau, Nardelli, Calandra, Car, Cavazzoni, Ceresoli,
  Cococcioni et~al.}}]{QE-2017}
\bibinfo{author}{\bibfnamefont{P.}~\bibnamefont{Giannozzi}},
  \bibinfo{author}{\bibfnamefont{O.}~\bibnamefont{Andreussi}},
  \bibinfo{author}{\bibfnamefont{T.}~\bibnamefont{Brumme}},
  \bibinfo{author}{\bibfnamefont{O.}~\bibnamefont{Bunau}},
  \bibinfo{author}{\bibfnamefont{M.~B.} \bibnamefont{Nardelli}},
  \bibinfo{author}{\bibfnamefont{M.}~\bibnamefont{Calandra}},
  \bibinfo{author}{\bibfnamefont{R.}~\bibnamefont{Car}},
  \bibinfo{author}{\bibfnamefont{C.}~\bibnamefont{Cavazzoni}},
  \bibinfo{author}{\bibfnamefont{D.}~\bibnamefont{Ceresoli}},
  \bibinfo{author}{\bibfnamefont{M.}~\bibnamefont{Cococcioni}},
  \bibnamefont{et~al.}, \bibinfo{journal}{Journal of Physics: Condensed Matter}
  \textbf{\bibinfo{volume}{29}}, \bibinfo{pages}{465901}
  (\bibinfo{year}{2017}),
  \urlprefix\url{http://stacks.iop.org/0953-8984/29/i=46/a=465901}.

\bibitem[{\citenamefont{Vanderbilt}(1990)}]{USPP_vanderbilt1990}
\bibinfo{author}{\bibfnamefont{D.}~\bibnamefont{Vanderbilt}},
  \bibinfo{journal}{Phys. Rev. B} \textbf{\bibinfo{volume}{41}},
  \bibinfo{pages}{7892} (\bibinfo{year}{1990}),
  \urlprefix\url{https://link.aps.org/doi/10.1103/PhysRevB.41.7892}.

\bibitem[{\citenamefont{Perdew and Yue}(1986)}]{GGA}
\bibinfo{author}{\bibfnamefont{J.~P.} \bibnamefont{Perdew}} \bibnamefont{and}
  \bibinfo{author}{\bibfnamefont{W.}~\bibnamefont{Yue}},
  \bibinfo{journal}{Physical Review B} \textbf{\bibinfo{volume}{33}},
  \bibinfo{pages}{8800} (\bibinfo{year}{1986}),
  \urlprefix\url{https://link.aps.org/doi/10.1103/PhysRevB.33.8800}.

\bibitem[{\citenamefont{Perdew et~al.}(1996)\citenamefont{Perdew, Burke, and
  Ernzerhof}}]{PBE}
\bibinfo{author}{\bibfnamefont{J.~P.} \bibnamefont{Perdew}},
  \bibinfo{author}{\bibfnamefont{K.}~\bibnamefont{Burke}}, \bibnamefont{and}
  \bibinfo{author}{\bibfnamefont{M.}~\bibnamefont{Ernzerhof}},
  \bibinfo{journal}{Physical Review Letters} \textbf{\bibinfo{volume}{77}},
  \bibinfo{pages}{3865} (\bibinfo{year}{1996}),
  \urlprefix\url{https://link.aps.org/doi/10.1103/PhysRevLett.77.3865}.

\bibitem[{\citenamefont{Cohen et~al.}(2008)\citenamefont{Cohen, Mori-Sánchez,
  and Yang}}]{PBE_bandgap_issue}
\bibinfo{author}{\bibfnamefont{A.~J.} \bibnamefont{Cohen}},
  \bibinfo{author}{\bibfnamefont{P.}~\bibnamefont{Mori-S\'{a}nchez}},
  \bibnamefont{and} \bibinfo{author}{\bibfnamefont{W.}~\bibnamefont{Yang}},
  \bibinfo{journal}{Science} \textbf{\bibinfo{volume}{321}},
  \bibinfo{pages}{792} (\bibinfo{year}{2008}),
  \urlprefix\url{http://www.jstor.org/stable/20144547}.

\bibitem[{\citenamefont{Heyd et~al.}(2003)\citenamefont{Heyd, Scuseria, and
  Ernzerhof}}]{HSE03}
\bibinfo{author}{\bibfnamefont{J.}~\bibnamefont{Heyd}},
  \bibinfo{author}{\bibfnamefont{G.~E.} \bibnamefont{Scuseria}},
  \bibnamefont{and}
  \bibinfo{author}{\bibfnamefont{M.}~\bibnamefont{Ernzerhof}},
  \bibinfo{journal}{The Journal of Chemical Physics}
  \textbf{\bibinfo{volume}{118}}, \bibinfo{pages}{8207} (\bibinfo{year}{2003}),
  \eprint{https://doi.org/10.1063/1.1564060}.

\bibitem[{\citenamefont{Heyd et~al.}(2006)\citenamefont{Heyd, Scuseria, and
  Ernzerhof}}]{HSE06}
\bibinfo{author}{\bibfnamefont{J.}~\bibnamefont{Heyd}},
  \bibinfo{author}{\bibfnamefont{G.~E.} \bibnamefont{Scuseria}},
  \bibnamefont{and}
  \bibinfo{author}{\bibfnamefont{M.}~\bibnamefont{Ernzerhof}},
  \bibinfo{journal}{The Journal of Chemical Physics}
  \textbf{\bibinfo{volume}{124}}, \bibinfo{pages}{219906}
  (\bibinfo{year}{2006}), \eprint{https://doi.org/10.1063/1.2204597}.

\bibitem[{\citenamefont{Noh et~al.}(2018)\citenamefont{Noh, Choi, Kim, Im, Kim,
  Seo, and Lee}}]{noh2018_PBE}
\bibinfo{author}{\bibfnamefont{G.}~\bibnamefont{Noh}},
  \bibinfo{author}{\bibfnamefont{D.}~\bibnamefont{Choi}},
  \bibinfo{author}{\bibfnamefont{J.-H.} \bibnamefont{Kim}},
  \bibinfo{author}{\bibfnamefont{D.-G.} \bibnamefont{Im}},
  \bibinfo{author}{\bibfnamefont{Y.-H.} \bibnamefont{Kim}},
  \bibinfo{author}{\bibfnamefont{H.}~\bibnamefont{Seo}}, \bibnamefont{and}
  \bibinfo{author}{\bibfnamefont{J.}~\bibnamefont{Lee}}, \bibinfo{journal}{Nano
  Letters} \textbf{\bibinfo{volume}{18}}, \bibinfo{pages}{4710}
  (\bibinfo{year}{2018}),
  \urlprefix\url{https://doi.org/10.1021/acs.nanolett.8b01030}.

\bibitem[{\citenamefont{Li et~al.}(2020)\citenamefont{Li, Chou, Hu, Plenio,
  Udvarhelyi, Thiering, Abdi, and Gali}}]{li2020_hse}
\bibinfo{author}{\bibfnamefont{S.}~\bibnamefont{Li}},
  \bibinfo{author}{\bibfnamefont{J.-P.} \bibnamefont{Chou}},
  \bibinfo{author}{\bibfnamefont{A.}~\bibnamefont{Hu}},
  \bibinfo{author}{\bibfnamefont{M.~B.} \bibnamefont{Plenio}},
  \bibinfo{author}{\bibfnamefont{P.}~\bibnamefont{Udvarhelyi}},
  \bibinfo{author}{\bibfnamefont{G.}~\bibnamefont{Thiering}},
  \bibinfo{author}{\bibfnamefont{M.}~\bibnamefont{Abdi}}, \bibnamefont{and}
  \bibinfo{author}{\bibfnamefont{A.}~\bibnamefont{Gali}}, \bibinfo{journal}{npj
  Quantum Information} \textbf{\bibinfo{volume}{6}}, \bibinfo{pages}{85}
  (\bibinfo{year}{2020}), ISSN \bibinfo{issn}{2056-6387},
  \urlprefix\url{https://doi.org/10.1038/s41534-020-00312-y}.

\bibitem[{\citenamefont{Monkhorst and Pack}(1976)}]{Monkhorst}
\bibinfo{author}{\bibfnamefont{H.~J.} \bibnamefont{Monkhorst}}
  \bibnamefont{and} \bibinfo{author}{\bibfnamefont{J.~D.} \bibnamefont{Pack}},
  \bibinfo{journal}{Physical Review B} \textbf{\bibinfo{volume}{13}},
  \bibinfo{pages}{5188} (\bibinfo{year}{1976}),
  \urlprefix\url{https://link.aps.org/doi/10.1103/PhysRevB.13.5188}.

\bibitem[{\citenamefont{Soykal et~al.}(2016)\citenamefont{Soykal, Dev, and
  Economou}}]{soykal2016silicon}
\bibinfo{author}{\bibfnamefont{{\"O}.}~\bibnamefont{Soykal}},
  \bibinfo{author}{\bibfnamefont{P.}~\bibnamefont{Dev}}, \bibnamefont{and}
  \bibinfo{author}{\bibfnamefont{S.~E.} \bibnamefont{Economou}},
  \bibinfo{journal}{Physical Review B} \textbf{\bibinfo{volume}{93}},
  \bibinfo{pages}{081207} (\bibinfo{year}{2016}),
  \urlprefix\url{https://doi.org/10.1103/PhysRevB.93.081207}.

\bibitem[{\citenamefont{Economou and Dev}(2016)}]{economou2016spin}
\bibinfo{author}{\bibfnamefont{S.~E.} \bibnamefont{Economou}} \bibnamefont{and}
  \bibinfo{author}{\bibfnamefont{P.}~\bibnamefont{Dev}},
  \bibinfo{journal}{Nanotechnology} \textbf{\bibinfo{volume}{27}},
  \bibinfo{pages}{504001} (\bibinfo{year}{2016}),
  \urlprefix\url{https://doi.org/10.1088/0957-4484/27/50/504001}.

\bibitem[{\citenamefont{Jia et~al.}(2020)\citenamefont{Jia, Gong, Li, Ma, Fang,
  Yang, and Cao}}]{passivated_SiC_NW_YaHui2020}
\bibinfo{author}{\bibfnamefont{Y.-H.} \bibnamefont{Jia}},
  \bibinfo{author}{\bibfnamefont{P.}~\bibnamefont{Gong}},
  \bibinfo{author}{\bibfnamefont{S.-L.} \bibnamefont{Li}},
  \bibinfo{author}{\bibfnamefont{W.-D.} \bibnamefont{Ma}},
  \bibinfo{author}{\bibfnamefont{X.-Y.} \bibnamefont{Fang}},
  \bibinfo{author}{\bibfnamefont{Y.-Y.} \bibnamefont{Yang}}, \bibnamefont{and}
  \bibinfo{author}{\bibfnamefont{M.-S.} \bibnamefont{Cao}},
  \bibinfo{journal}{Physics Letters A} \textbf{\bibinfo{volume}{384}},
  \bibinfo{pages}{126106} (\bibinfo{year}{2020}), ISSN
  \bibinfo{issn}{0375-9601},
  \urlprefix\url{https://www.sciencedirect.com/science/article/pii/S0375960119310151}.

\bibitem[{\citenamefont{Tanaka et~al.}(2003)\citenamefont{Tanaka, Matsunaga,
  Ikuhara, and Yamamoto}}]{tanaka2003}
\bibinfo{author}{\bibfnamefont{T.}~\bibnamefont{Tanaka}},
  \bibinfo{author}{\bibfnamefont{K.}~\bibnamefont{Matsunaga}},
  \bibinfo{author}{\bibfnamefont{Y.}~\bibnamefont{Ikuhara}}, \bibnamefont{and}
  \bibinfo{author}{\bibfnamefont{T.}~\bibnamefont{Yamamoto}},
  \bibinfo{journal}{Phys. Rev. B} \textbf{\bibinfo{volume}{68}},
  \bibinfo{pages}{205213} (\bibinfo{year}{2003}),
  \urlprefix\url{https://link.aps.org/doi/10.1103/PhysRevB.68.205213}.

\bibitem[{\citenamefont{Dev}(2020)}]{PDEV_hBN_2020}
\bibinfo{author}{\bibfnamefont{P.}~\bibnamefont{Dev}},
  \bibinfo{journal}{Physical Review Research} \textbf{\bibinfo{volume}{2}},
  \bibinfo{pages}{022050(R)} (\bibinfo{year}{2020}),
  \urlprefix\url{https://link.aps.org/doi/10.1103/PhysRevResearch.2.022050}.

\bibitem[{\citenamefont{Narayanan and Dev}(2023)}]{narayanan2023}
\bibinfo{author}{\bibfnamefont{S.~K.} \bibnamefont{Narayanan}}
  \bibnamefont{and} \bibinfo{author}{\bibfnamefont{P.}~\bibnamefont{Dev}},
  \bibinfo{journal}{ACS Applied Nano Materials} \textbf{\bibinfo{volume}{6}},
  \bibinfo{pages}{3446} (\bibinfo{year}{2023}),
  \urlprefix\url{https://doi.org/10.1021/acsanm.2c05233}.

\bibitem[{\citenamefont{Torpo et~al.}(1999)\citenamefont{Torpo, Nieminen,
  Laasonen, and Pöykkö}}]{VSiq0_Spin1_Torpo1999}
\bibinfo{author}{\bibfnamefont{L.}~\bibnamefont{Torpo}},
  \bibinfo{author}{\bibfnamefont{R.~M.} \bibnamefont{Nieminen}},
  \bibinfo{author}{\bibfnamefont{K.~E.} \bibnamefont{Laasonen}},
  \bibnamefont{and} \bibinfo{author}{\bibfnamefont{S.}~\bibnamefont{P\"{o}ykk\"{o}}},
  \bibinfo{journal}{Applied Physics Letters} \textbf{\bibinfo{volume}{74}},
  \bibinfo{pages}{221} (\bibinfo{year}{1999}),
  \urlprefix\url{https://doi.org/10.1063/1.123299}.

\bibitem[{\citenamefont{Mizuochi et~al.}(2002)\citenamefont{Mizuochi, Yamasaki,
  Takizawa, Morishita, Ohshima, Itoh, and Isoya}}]{Mizuochi2002}
\bibinfo{author}{\bibfnamefont{N.}~\bibnamefont{Mizuochi}},
  \bibinfo{author}{\bibfnamefont{S.}~\bibnamefont{Yamasaki}},
  \bibinfo{author}{\bibfnamefont{H.}~\bibnamefont{Takizawa}},
  \bibinfo{author}{\bibfnamefont{N.}~\bibnamefont{Morishita}},
  \bibinfo{author}{\bibfnamefont{T.}~\bibnamefont{Ohshima}},
  \bibinfo{author}{\bibfnamefont{H.}~\bibnamefont{Itoh}}, \bibnamefont{and}
  \bibinfo{author}{\bibfnamefont{J.}~\bibnamefont{Isoya}},
  \bibinfo{journal}{Phys. Rev. B} \textbf{\bibinfo{volume}{66}},
  \bibinfo{pages}{235202} (\bibinfo{year}{2002}),
  \urlprefix\url{https://link.aps.org/doi/10.1103/PhysRevB.66.235202}.

\bibitem[{\citenamefont{Janz{\'e}n et~al.}(2009)\citenamefont{Janz{\'e}n, Gali,
  Carlsson, G{\"a}llstr{\"o}m, Magnusson, and Son}}]{janzen2009silicon}
\bibinfo{author}{\bibfnamefont{E.}~\bibnamefont{Janz{\'e}n}},
  \bibinfo{author}{\bibfnamefont{A.}~\bibnamefont{Gali}},
  \bibinfo{author}{\bibfnamefont{P.}~\bibnamefont{Carlsson}},
  \bibinfo{author}{\bibfnamefont{A.}~\bibnamefont{G{\"a}llstr{\"o}m}},
  \bibinfo{author}{\bibfnamefont{B.}~\bibnamefont{Magnusson}},
  \bibnamefont{and} \bibinfo{author}{\bibfnamefont{N.~T.} \bibnamefont{Son}},
  \bibinfo{journal}{Physica B: Condensed Matter}
  \textbf{\bibinfo{volume}{404}}, \bibinfo{pages}{4354} (\bibinfo{year}{2009}).

\bibitem[{\citenamefont{Dev et~al.}(2008)\citenamefont{Dev, Xue, and
  Zhang}}]{Dev_PRL_DeepDefects_2008}
\bibinfo{author}{\bibfnamefont{P.}~\bibnamefont{Dev}},
  \bibinfo{author}{\bibfnamefont{Y.}~\bibnamefont{Xue}}, \bibnamefont{and}
  \bibinfo{author}{\bibfnamefont{P.}~\bibnamefont{Zhang}},
  \bibinfo{journal}{Physical Review Letters} \textbf{\bibinfo{volume}{100}},
  \bibinfo{pages}{117204} (\bibinfo{year}{2008}),
  \urlprefix\url{https://link.aps.org/doi/10.1103/PhysRevLett.100.117204}.

\bibitem[{\citenamefont{Dev and Zhang}(2010)}]{Dev_PRB_DeepDefects_2010}
\bibinfo{author}{\bibfnamefont{P.}~\bibnamefont{Dev}} \bibnamefont{and}
  \bibinfo{author}{\bibfnamefont{P.}~\bibnamefont{Zhang}},
  \bibinfo{journal}{Physical Review B} \textbf{\bibinfo{volume}{81}},
  \bibinfo{pages}{085207} (\bibinfo{year}{2010}),
  \urlprefix\url{https://link.aps.org/doi/10.1103/PhysRevB.81.085207}.

\bibitem[{\citenamefont{Dev et~al.}(2010)\citenamefont{Dev, Zeng, and
  Zhang}}]{Dev_PRB_NW_2010}
\bibinfo{author}{\bibfnamefont{P.}~\bibnamefont{Dev}},
  \bibinfo{author}{\bibfnamefont{H.}~\bibnamefont{Zeng}}, \bibnamefont{and}
  \bibinfo{author}{\bibfnamefont{P.}~\bibnamefont{Zhang}},
  \bibinfo{journal}{Physical Review B} \textbf{\bibinfo{volume}{82}},
  \bibinfo{pages}{165319} (\bibinfo{year}{2010}),
  \urlprefix\url{https://link.aps.org/doi/10.1103/PhysRevB.82.165319}.

\bibitem[{\citenamefont{Huang et~al.}(2008)\citenamefont{Huang, Sun, Tao,
  Menard, Nuzzo, and Zuo}}]{Huang2008SkinEffect}
\bibinfo{author}{\bibfnamefont{W.~J.} \bibnamefont{Huang}},
  \bibinfo{author}{\bibfnamefont{R.}~\bibnamefont{Sun}},
  \bibinfo{author}{\bibfnamefont{J.}~\bibnamefont{Tao}},
  \bibinfo{author}{\bibfnamefont{L.~D.} \bibnamefont{Menard}},
  \bibinfo{author}{\bibfnamefont{R.~G.} \bibnamefont{Nuzzo}}, \bibnamefont{and}
  \bibinfo{author}{\bibfnamefont{J.~M.} \bibnamefont{Zuo}},
  \bibinfo{journal}{Nature Materials} \textbf{\bibinfo{volume}{7}},
  \bibinfo{pages}{308} (\bibinfo{year}{2008}),
  \urlprefix\url{https://doi.org/10.1038/nmat2132}.

\bibitem[{\citenamefont{Gamble}(1974)}]{Gamble1974}
\bibinfo{author}{\bibfnamefont{F.}~\bibnamefont{Gamble}},
  \bibinfo{journal}{Journal of Solid State Chemistry}
  \textbf{\bibinfo{volume}{9}}, \bibinfo{pages}{358} (\bibinfo{year}{1974}),
  ISSN \bibinfo{issn}{0022-4596},
  \urlprefix\url{https://www.sciencedirect.com/science/article/pii/0022459674900954}.

\bibitem[{\citenamefont{Kumar et~al.}(2022)\citenamefont{Kumar, Sharma,
  Shirodkar, and Dev}}]{ML_Pankaj2022}
\bibinfo{author}{\bibfnamefont{P.}~\bibnamefont{Kumar}},
  \bibinfo{author}{\bibfnamefont{V.}~\bibnamefont{Sharma}},
  \bibinfo{author}{\bibfnamefont{S.~N.} \bibnamefont{Shirodkar}},
  \bibnamefont{and} \bibinfo{author}{\bibfnamefont{P.}~\bibnamefont{Dev}},
  \bibinfo{journal}{Phys. Rev. Mater.} \textbf{\bibinfo{volume}{6}},
  \bibinfo{pages}{094007} (\bibinfo{year}{2022}),
  \urlprefix\url{https://link.aps.org/doi/10.1103/PhysRevMaterials.6.094007}.

\bibitem[{\citenamefont{Sz\'asz et~al.}(2015)\citenamefont{Sz\'asz, Iv\'ady,
  Abrikosov, Janz\'en, Bockstedte, and Gali}}]{szasz2015}
\bibinfo{author}{\bibfnamefont{K.}~\bibnamefont{Sz\'asz}},
  \bibinfo{author}{\bibfnamefont{V.}~\bibnamefont{Iv\'ady}},
  \bibinfo{author}{\bibfnamefont{I.~A.} \bibnamefont{Abrikosov}},
  \bibinfo{author}{\bibfnamefont{E.}~\bibnamefont{Janz\'en}},
  \bibinfo{author}{\bibfnamefont{M.}~\bibnamefont{Bockstedte}},
  \bibnamefont{and} \bibinfo{author}{\bibfnamefont{A.}~\bibnamefont{Gali}},
  \bibinfo{journal}{Phys. Rev. B} \textbf{\bibinfo{volume}{91}},
  \bibinfo{pages}{121201} (\bibinfo{year}{2015}),
  \urlprefix\url{https://link.aps.org/doi/10.1103/PhysRevB.91.121201}.

\bibitem[{\citenamefont{Gordon et~al.}(2015)\citenamefont{Gordon, Janotti, and
  Van~de Walle}}]{gordon2015_VSim1_HSE}
\bibinfo{author}{\bibfnamefont{L.}~\bibnamefont{Gordon}},
  \bibinfo{author}{\bibfnamefont{A.}~\bibnamefont{Janotti}}, \bibnamefont{and}
  \bibinfo{author}{\bibfnamefont{C.~G.} \bibnamefont{Van~de Walle}},
  \bibinfo{journal}{Phys. Rev. B} \textbf{\bibinfo{volume}{92}},
  \bibinfo{pages}{045208} (\bibinfo{year}{2015}),
  \urlprefix\url{https://link.aps.org/doi/10.1103/PhysRevB.92.045208}.

\end{thebibliography}

\end{document}